\documentclass[letterpaper]{article} \usepackage{etoolbox}
\providetoggle{submission}
\providetoggle{aaai}
\providetoggle{arxiv}
\settoggle{submission}{true}  \settoggle{aaai}{false}        \settoggle{arxiv}{true}      

\iftoggle{submission}{
  \iftoggle{aaai}{
    \usepackage[submission]{aaai2026}  }{
    \usepackage{aaai2026}  }
}{
  \usepackage{aaai2026}  }

\iftoggle{arxiv}{
  \nocopyright
}{}

\usepackage{times}  \usepackage{helvet}  \usepackage{courier}  \usepackage{xcolor}
\usepackage{longtable}
\usepackage[hyphens]{url}  \usepackage{graphicx}    \usepackage{natbib}  \usepackage{soul}
\usepackage{booktabs}
\usepackage{caption}     \usepackage{algorithm}
\usepackage{algorithmic}
\usepackage{tcolorbox}
\usepackage{natbib}
\usepackage{amsmath}
\usepackage{amssymb}
\usepackage{array}
\usepackage{enumitem}
\usepackage{ifthen}

\newcommand{\spheading}[2][5em]{\rotatebox{70}{\parbox{#1}{\raggedright #2}}}

\newcommand{\hospital}[1]{\ifthenelse{\equal{#1}{anonymous}}{Anonymous Hospital}{Zuckerberg San Francisco General Hospital}}

\newcommand{\repourl}[1]{\ifthenelse{\equal{#1}{anonymous}}{https://anonymous.4open.science/r/social-wayfinder-7E0B}{https://github.com/jjfenglab/social-wayfinder}}
\newcommand{\numattributes}{15 }
\newcommand{\promptone}{\textit{Prompt1}}
\newcommand{\prompttwo}{\textit{Prompt2}}
\newcommand{\promptthree}{\textit{Prompt3}}
\newcommand{\promptfour}{\textit{Prompt4}}

\usepackage{newfloat}
\newcolumntype{H}{>{\setbox0=\hbox\bgroup}c<{\egroup}@{}}
\usepackage{listings}
\DeclareCaptionStyle{ruled}{labelfont=normalfont,labelsep=colon,strut=off} 
\floatstyle{ruled}
\newfloat{listing}{tb}{lst}{}
\floatname{listing}{Listing}
\title{
When the Domain Expert Has No Time and the LLM Developer Has No Clinical Expertise: Real-World Lessons from LLM Co-Design in a Safety-Net Hospital
}
\iftoggle{submission}{
  \iftoggle{aaai}{
    \author{Anonymous}
    \affiliations{Anonymous}
  }{
    \author{
        Avni Kothari\iftoggle{submission}{\iftoggle{aaai}{}{\footnote{For correspondence  please contact: avni.kothari@ucsf.edu}}}{\footnote{For correspondence  please contact: avni.kothari@ucsf.edu}}\textsuperscript{1, 2}, Patrick Vossler\textsuperscript{1, 2}, Jean Digitale\textsuperscript{1}, Mohammad Forouzannia\textsuperscript{1}, Elise Rosenberg\textsuperscript{2}, Michele Lee\textsuperscript{2}, Jenneé Bryant\textsuperscript{2}, Melanie Molina\textsuperscript{1, 2}, James Marks\textsuperscript{1, 2}, Lucas Zier\textsuperscript{1, 2}, Jean Feng\textsuperscript{1, 2}
    }
    \affiliations{
        \textsuperscript{1}University of California, San Francisco\\
        \textsuperscript{2}Zuckerberg San Francisco General Hospital
    }
  }
}{
  \author{
      Avni Kothari\textsuperscript{1, 2}, Patrick Vossler\textsuperscript{1, 2}, Jean Digitale\textsuperscript{2}, Mohammad Forouzannia\textsuperscript{2}, Elise Rosenberg\textsuperscript{2}, Michele Lee\textsuperscript{2}, Jenneé Bryant\textsuperscript{2}, Melanie Molina\textsuperscript{2}, James Marks\textsuperscript{2}, Lucas Zier\textsuperscript{2}, Jean Feng\textsuperscript{1}
  }
  \affiliations{
      \textsuperscript{1}University of California, San Francisco\\
      \textsuperscript{2}Zuckerberg San Francisco General Hospital\\
  }
}

\begin{document}

\maketitle

\begin{abstract}
Large language models (LLMs) have the potential to address social and behavioral determinants of health by transforming labor intensive workflows in resource-constrained settings. Creating LLM-based applications that serve the needs of underserved communities requires a deep understanding of their local context, but it is often the case that neither LLMs nor their developers possess this local expertise, and the experts in these communities often face severe time/resource constraints.
This creates a disconnect: how can one engage in meaningful co-design of an LLM-based application for an under-resourced community when the communication channel between the LLM developer and domain expert is constrained? 
We explored this question through a real-world case study, in which our data science team sought to partner with social workers at a safety net hospital to build an LLM application that summarizes patients' social needs. 
Whereas prior works focus on the challenge of prompt tuning, we found that the most critical challenge in this setting is the careful and precise specification of \textit{what} information to surface to providers so that the LLM application is accurate, comprehensive, and verifiable.
Here we present a novel co-design framework for settings with limited access to domain experts, in which the summary generation task is first decomposed into individually-optimizable attributes and then each attribute is efficiently refined and validated through a multi-tier cascading approach.
\end{abstract}

\section{Introduction}

Safety-net programs deliver essential medical and social services to underserved communities, yet they often operate under severe resource shortages and staffing constraints. Recent advances in LLMs offer tremendous promise for amplifying services provided by these programs, but realizing that potential requires genuine participatory design. Without domain expertise and stakeholder input, AI solutions are at a substantially higher risk of perpetuating biases, being misaligned, or lacking practical utility \citep{de2020case, buolamwini2018gender, shneiderman2020human}.
However, unlike standard software engineering, co-designing an LLM application requires overcoming the ``gulf of envisionment'' \citep{Subramonyam2024-at}: there is often a large gap between a human's initial intentions to use an LLM and their final crystallized intentions that are translatable into effective LLM prompts.
Crossing this gulf can often be a long process, because the exact intentions of a project as well as an LLM's ability to produce the desired outputs are typically unclear in the initial stages.

\begin{figure}
    \centering
    \includegraphics[width=0.9\linewidth]{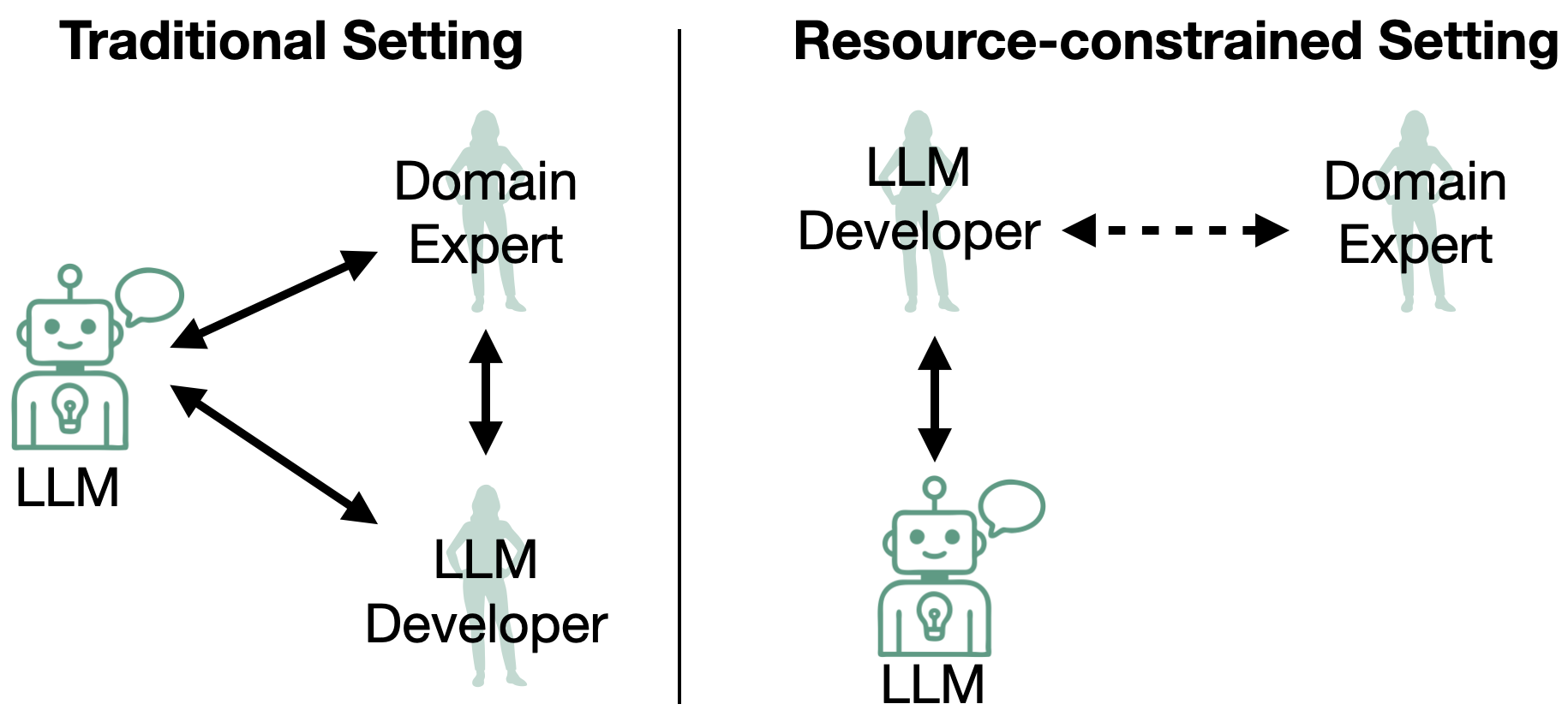}
    \caption{
    In traditional settings, both the LLM developer and domain expert have frequent communication to jointly iterate on and create an LLM application (solid lines mean full access). By contrast, in resource-constrained settings, the LLM developer has limited access (dashed lines) to the domain expert, which can be a major barrier to developing a truly useful application.
    }
    \label{fig:pipeline}
    \vspace{-0.5cm}
\end{figure}

Traditional approaches to LLM application co-design require significant time investment from both LLM developers and domain experts, as they assume the two parties will work closely to iteratively refine their intentions and prompts \citep{karayanni2024keeping}.
However, this overlooks a major catch-22 when developing AI for resource-constrained settings (Fig~\ref{fig:pipeline}):
the very communities that could benefit from AI assistance are also those whose resource constraints prevent them from fully engaging in the intensive co-design process.
Without the help of domain experts, AI developers may not have the domain expertise and/or the full local context to build a truly useful application.

We examined challenges in LLM co-design in under-resourced settings through the lens of a real-world case study, in which our data science team was asked by the local safety-net hospital to build an LLM application to assist the hospital's inpatient social workers (SWs).
The primary responsibility of SWs---the domain experts in this setting---is to address social and behavioral determinants of health (SBDH) needs of patients and facilitate their safe discharge.
The specific aim was to help with the ``pre-chart summarization'' step of the SW workflow, in which they identify a patient's SBDH across various domains, such as food, housing, and transportation, by reviewing past patient records.
This step is necessary to prepare SWs for in-person discussions with the patient and to locate suitable social services.
It is also extremely time-intensive, requiring manual review of fragmented information across multiple encounters in patient charts. A team of 23 inpatient SWs at the hospital collectively spends more than 300 hours per week on pre-chart summarization across their caseloads. 
This is problematic in a safety-net hospital that faces resource constraints and understaffing, where the number of patients who need social services exceeds the team's current capacity.
LLMs have the potential to substantially accelerate the SWs' workflow by extracting and summarizing SBDH information from clinical notes.

Nevertheless, our data science team met major practical barriers in our initial attempts to use existing approaches.
The chief challenge is that much of the existing LLM literature assumes the intention, i.e. the desired goals for using an LLM, are already precisely specified.
However, unlike well-studied AI applications, novel uses in resource-constrained settings often lack established patterns or precedents, forcing teams to simultaneously define the problem space and develop solutions.
Operating alone, neither the AI developer nor the domain expert was able to fully crystallize their intentions, as the AI developer has limited domain expertise and the resource-constrained domain expert has limited bandwidth.
As such, methods such as human-driven prompt tuning \cite{goyal2022news, zhang2024benchmarking, aly2025evaluation, williams2025physician, williams2025evaluating} and auto-prompting \cite{Arawjo_2024, kim2024evallm, shankar2024validates, he2025crispo, Wang2024-zc} lack sufficient utility and/or applicability in resource-constrained settings.
Furthermore, existing co-design frameworks were similarly impractical, as the co-design process was too burdensome for experts from these communities \cite{majumdar2025hardness}.
Finally, beyond intention formation, LLM validation was also challenging for the very same reasons.

\begin{figure}
    \centering
\includegraphics[width=\linewidth]{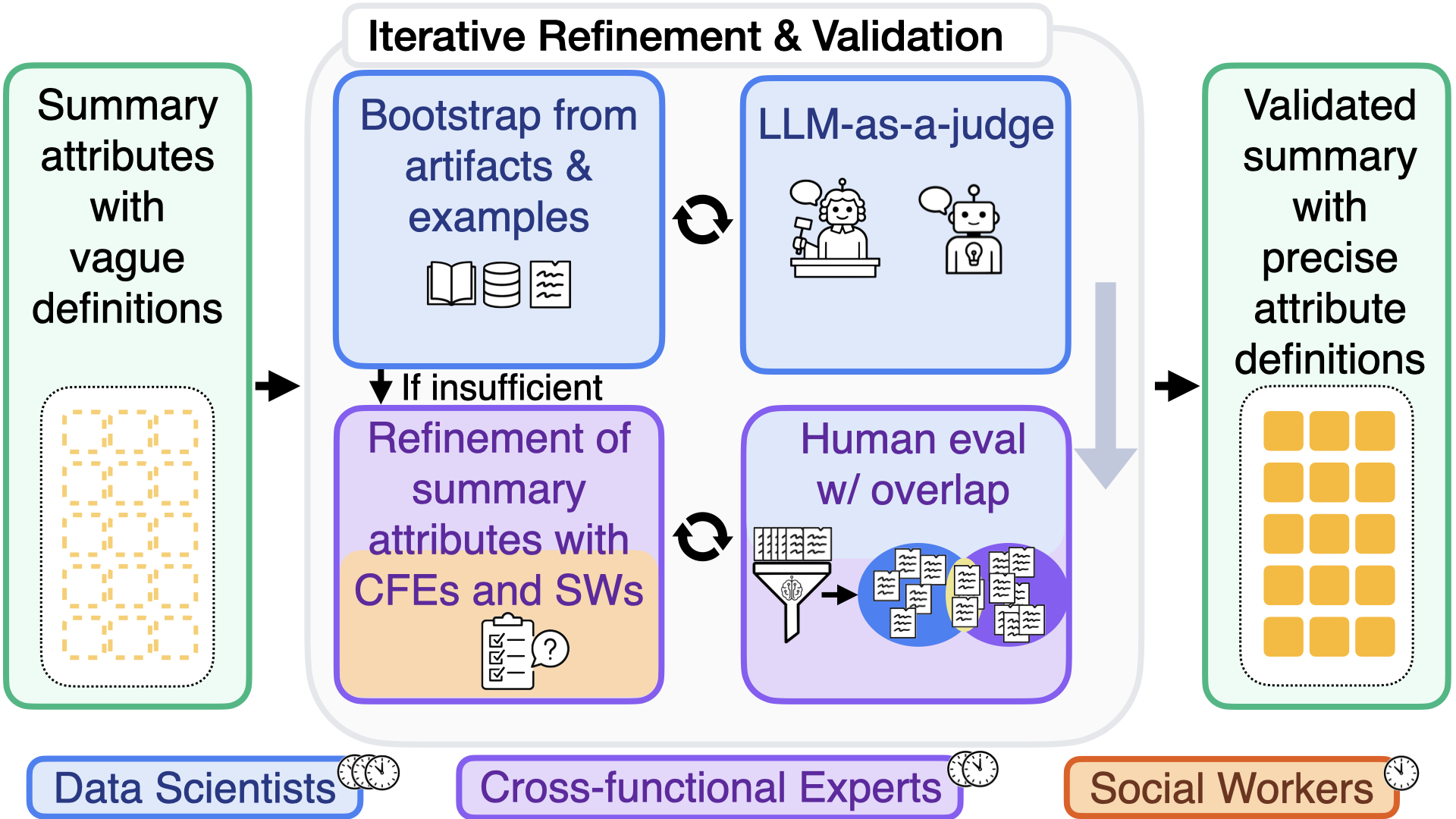}
    \caption{To efficiently co-design an LLM application starting from vaguely defined project goals, the framework recruits cross-functional experts (CFEs) to close the expertise gap, decomposes the task of generating free-form summaries into structured attributes, and iteratively refines and validates each attribute. A multi-tier cascading approach is used to minimize the number of requests sent to the domain experts who had limited availability (social workers (SWs) in this case study).
    }
    \label{fig:framework-fig}
\end{figure}

Overcoming these barriers requires new design strategies.
We introduce a dynamic, multi-tier framework that was significantly more effective and efficient in leveraging our resource-constrained domain experts through the following strategies (Fig~\ref{fig:framework-fig}):
\begin{itemize}
    \item \ul{Assemble a team to close the gap}: Assemble a team that fills the gap in domain expertise and availability, so that the team is not just at the extreme ends of the spectrum.
    \item \ul{Decomposing free-form summaries into attributes}: Modularize the summarization task into semi-structured attributes that can be individually optimized using fragmented data and targeted feedback.
    \item \ul{Multi-tier cascade for intention formation}: 
    Iteratively refine the precise definition of each summary attribute by bootstrapping from existing organizational artifacts and then dynamically eliciting targeted, concrete feedback through a multi-tier cascade.
    \item \ul{Multi-tier validation}: Validate the extracted summary attributes using a multi-tier cascade, in which LLM-as-a-judge serves as a scalable but imperfect guide and human experts provide higher-quality targeted feedback.
\end{itemize}
Through this approach, the first prototype of the LLM application achieved high overall accuracy and was viewed by SWs at the safety net hospital as being very likely to accelerate their workflow.

\section{Problem Definition}
Our aim was to develop an LLM-based summarization tool for SWs at our local safety-net hospital that was accurate, comprehensive, useful, and verifiable.
Pre-chart summarization requires synthesizing information from multiple note types in the electronic health record (EHR), including History \& Physical examinations, discharge summaries, progress notes, and various specialty consults spanning psychiatry, social work, behavioral health, care management teams across inpatient, outpatient, and emergency department contexts.
Key SBDH domains to be covered included housing stability, food security, substance use, mental health, safety concerns, insurance status, immigration-related barriers, transportation access, durable medical equipment needs, and outpatient therapy arrangements.

While the task initially appeared well-scoped---synthesize social needs from EHR notes across defined domains---it presented four obstacles that are common when developing AI for under-resourced settings. 
Here we detail how each challenge manifested in our specific context and how it inhibited the use of existing methods:

\textbf{(C1) Resource constraints and expertise gap:}
Despite having close collaborations with the safety net hospital, the data science team had limited expertise in social work and thus did not have the skillset needed to craft a high-quality prompt alone.
Simultaneously, the SWs, managing large caseloads, could only commit to meeting one hour once every few weeks, with occasional emails in between.

\textbf{(C2) Lack of gold-standard summaries:}
Pre-chart summarization existed entirely as tacit knowledge. It was an undocumented process that happens primarily ``in the SW's head'' with no written examples, templates, or formal documentation.
Without existing summaries, we could not fine-tune or train LLMs for this specific task. 
Similarly, the absence of gold-standard labeled data made automatic prompt tuning methods infeasible, as there was no consensus on what constituted a good summary.

\textbf{(C3) Underspecified design requirements:}
Neither existing literature nor institutional materials provided templates for pre-chart summarization, particularly for safety-net populations.
Beyond lacking prompting guidance, we faced a more fundamental challenge: determining which of the many details in a patient's EHR were relevant to surface in the LLM application. The specification task of defining what constituted a useful summary required domain expertise that the data science team lacked.

\textbf{(C4): Evaluation:} Rigorous validation of LLM extractions is necessary for safe and effective us of AI in high-stakes domains like healthcare. Due to our limited resources, a key concern was balancing the accuracy and reliability of the LLM application with resource use/cost.

\section{Related Work}
\textbf{LLM summarization in healthcare}: Prior works have demonstrated that LLMs have the potential to generate useful summaries of patient charts \citep{williams2025physician, williams2025evaluating, bednarczyk2025scientific, Gero2024-bn}.
However, evaluating these summaries is challenging \cite{alaaposition}, as they rely on human effort to track different attributes and facts.
Closest to this work, \cite{Gero2024-bn} proposes structuring summaries into individual attributes to assist with LLM-as-a-judge evaluations.
However, \cite{Gero2024-bn} uses vaguely-defined attributes from a general ontology.
As such, their LLM-as-a-judge pipeline finds that the generated extractions have moderate agreement rates with the ground truth.
In contrast, this work collaborates with domain experts to construct attributes that are precise and relevant to local needs, resulting in much higher agreement rates.

\textbf{LLM Development with Human in the loop}:
Prior methods to develop LLM based applications focus on collaborating with the domain expert to iteratively co-design a product. This is done through gathering product requirements, iterating on the prompt \citep{reza2025prompthive, karayanni2024keeping, Shah2023-tt} and assessing the evaluation criteria through tools such as EvalLM \citep{kim2024evallm}, EvalGen \citep{shankar2024validates}, ChainForge  \citep{Arawjo_2024}, and LLM Comparator \citep{kahng2024llmcomparatorvisualanalytics}. These frameworks develop user interfaces for domain experts to interact with to iteratively improve the LLM system. However, these approaches do not address low-resource settings where domain experts have limited availability and would require significant time to learn the terminology and tools used in LLM applications. Our framework aimed to address the challenge of incorporating domain expertise despite this challenge.

\textbf{Autoprompting}:
Auto-prompting methods automatically improve the prompt by providing feedback to an LLM \citep{shin2020autoprompt, yang2023large, zhou2022large, pryzant2023automatic}. Numerous tools such as DsPy \citep{khattab2023dspy}, Textgrad \citep{yuksekgonul2024textgrad}, and Claude's prompt optimizer have been developed to help find the most optimal prompts. Generally, these tools require a ground truth dataset for optimization, but recent methods have proposed solutions for optimizing without a labeled evaluation dataset, a regime we were currently working under. These approaches structure evaluation by either asking a domain expert or DS \cite{Arawjo_2024, kim2024evallm, shankar2024validates, he2025crispo, Wang2024-zc} to provide feedback on the outputs by using multiple LLM judges to critique the LLM candidate's answer \citep{zhu2023judgelm, zheng2023judging}. When domain experts are asked to provide feedback, they still require significant time from them to understand LLM basics, learn how to provide feedback, and annotate a large number of examples across all the prompt iterations \citep{szymanski2024comparing, karayanni2024keeping}. On the other hand, while using multiple LLM judges can reduce expert burden, this approach often fails to clarify domain-specific requirements and definitions. This has made autoprompting most successful for optimizing over a predefined evaluation metric rather than precisely defining domain-specific requirements. Our framework addressed scenarios where we had to simultaneously optimize a prompt and define product requirements.

\textbf{Clinical Social Needs Extractions}:
Prior methods for social needs extraction from clinical notes focus on the setting of single notes that are brief and well-structured, whereas real-world settings involve complex, longitudinal notes \citep{Yu2024-he, Guevara2024-ul, lybarger2023leveraging, mahbub2024leveraging, Bedi2025-ec}.
Furthermore, clinical notes for safety-net patients are incredibly more complex, as any single note may contain numerous references to SBDH, as well as SW jargon and abbreviations.
Finally, existing methods employ generic definitions of social needs that fail to capture local contexts and practical utility in day-to-day settings, and do not address a key responsibility of clinical SWs: facilitating safe patient discharge. In contrast, our aim was to develop an LLM application that addresses the real-world needs of SWs.

\section{Initial (Failed) Attempts}

We initially tried applying existing LLM application design techniques to this problem setting, a subset of which are listed in Table~\ref{tab:failed_attempts}.
These initial attempts failed because existing methods were met with various barriers in this resource-constrained setting.
While the attempts were individually unsuccessful, they highlighted two critical learnings that motivated our eventual design strategy.

\begin{table}
    \centering
    \small
    \begin{tabular}{p{2.5cm}|p{1.5cm}|p{3.5cm}}
        \textbf{Approach} & \textbf{Challenge} & \textbf{Failure Modes} \\
        \toprule
        DSs shadow SWs due to time constraints, then independently author prompts & (C1) Resource constraints and expertise gap & DSs still lack sufficient understanding of the practical, real-world needs that SW encounter \\
        \midrule
        Use previous social work consult notes as ground truth summaries & (C2) Lack of gold standard summaries & Past notes are a reflection of verbal conversations with the patient, rather than summaries of the patient record
        \\
        \midrule
        SWs specify general requirements, DSs translate specifications into a prompt, and SWs are asked to prompt tune &
        (C3) Underspecified design requirements & DSs could not translate vague requirements into clear prompts, and SWs had insufficient bandwidth to meaningfully iterate on prompt tuning for complex summaries
         \\
        \midrule
        Ask SWs to annotate clinical notes for relevant SBDH information & (C4) Evaluation & SWs had time to annotate a few notes at most, which is insufficient for comprehensive evaluation  \\
        \bottomrule
    \end{tabular}
    \caption{Initial attempts to develop an LLM application in resource-constrained setting using existing approaches. DS=data scientist, SW=social worker}
    \label{tab:failed_attempts}
    \vspace{-0.2cm}
\end{table}

First, it was evident that the DSs could not rely heavily on the SWs to define the AI application or guide the AI model itself.
There needed to be a way to ``bootstrap'' the process without needing SW input (as in ``pull oneself up by one's bootstraps''), so that SWs could provide meaningful feedback with the limited bandwidth they had.
This meant finding existing organizational resources that could help clarify what information to extract using the LLM and adding new team members who could help bridge the large gap between the DSs and SWs.

Second, the goal of generating completely free-form AI-written summaries was too ill-defined.
SWs found that requests for well-specified templates and examples were too time-consuming. Furthermore, free-form summaries are too difficult to evaluate objectively. Instead, we needed to break down a summary into individual attributes. Through modularization, we could individually refine each attribute's precise definition and prompt-tune in a much more targeted and efficient fashion.

\section{Resource-efficient LLM  Co-Design}
To conduct LLM application co-design in this resource-constrained setting, we introduced a modular, multi-tiered approach that allows different contributors to focus on individual components and iterative refinement across tiers (Fig~\ref{fig:framework-fig}).
This began with organizing a team with a sufficient range of expertise and availability (Step 1) and decomposing our complex task into manageable components (Step 2) to facilitate better collaboration and prompt refinement.
The LLM application was then iteratively refined by cycling between tiered tuning of the attribute definition and prompt (Step 3) and tiered validation of the LLM extractions (Step 4).
We describe the general design strategies and then how each were implemented in this case study.

\subsection{Step 1: Assemble a team that closes the gap}

\begin{tcolorbox}[boxsep=0pt, left=3pt, right=3pt, top=2pt]
\textbf{Design strategy}: 
Traditional LLM application development relies solely on domain experts and data scientists, which creates a bottleneck in resource-constrained settings. To fill this gap, include additional team members with cross-functional expertise who have sufficient availability to serve as the bridge and buffer between DSs and the full domain experts.
\end{tcolorbox}

\textbf{Implementation}: 
Our team was initially composed of those at the extreme ends of the domain expertise and availability spectrum, with DSs at one end and SWs at the other.
To fill the gap, we recruited cross-functional experts (CFEs) who had moderate, though not full, availability and interdisciplinary expertise spanning real-world clinical work and data science.
Thus the final team consisted of SWs who have deep subject matter knowledge but were limited to only a few hours of contribution per month, CFEs who had sufficient time and context to help translate some of the SW needs to DSs, and the DSs who were the most available but lacked domain expertise.
(More specifically, our CFE team members consisted of clinical care providers and clinical researchers with some data science exposure.)

By having a range in expertise and availability levels, the DSs would escalate questions to SWs only when necessary, ensuring that each group's time was used as efficiently as possible. 

\subsection{Step 2: Decompose the free-form summary}

\begin{figure}
    \centering
    \includegraphics[width=\linewidth]{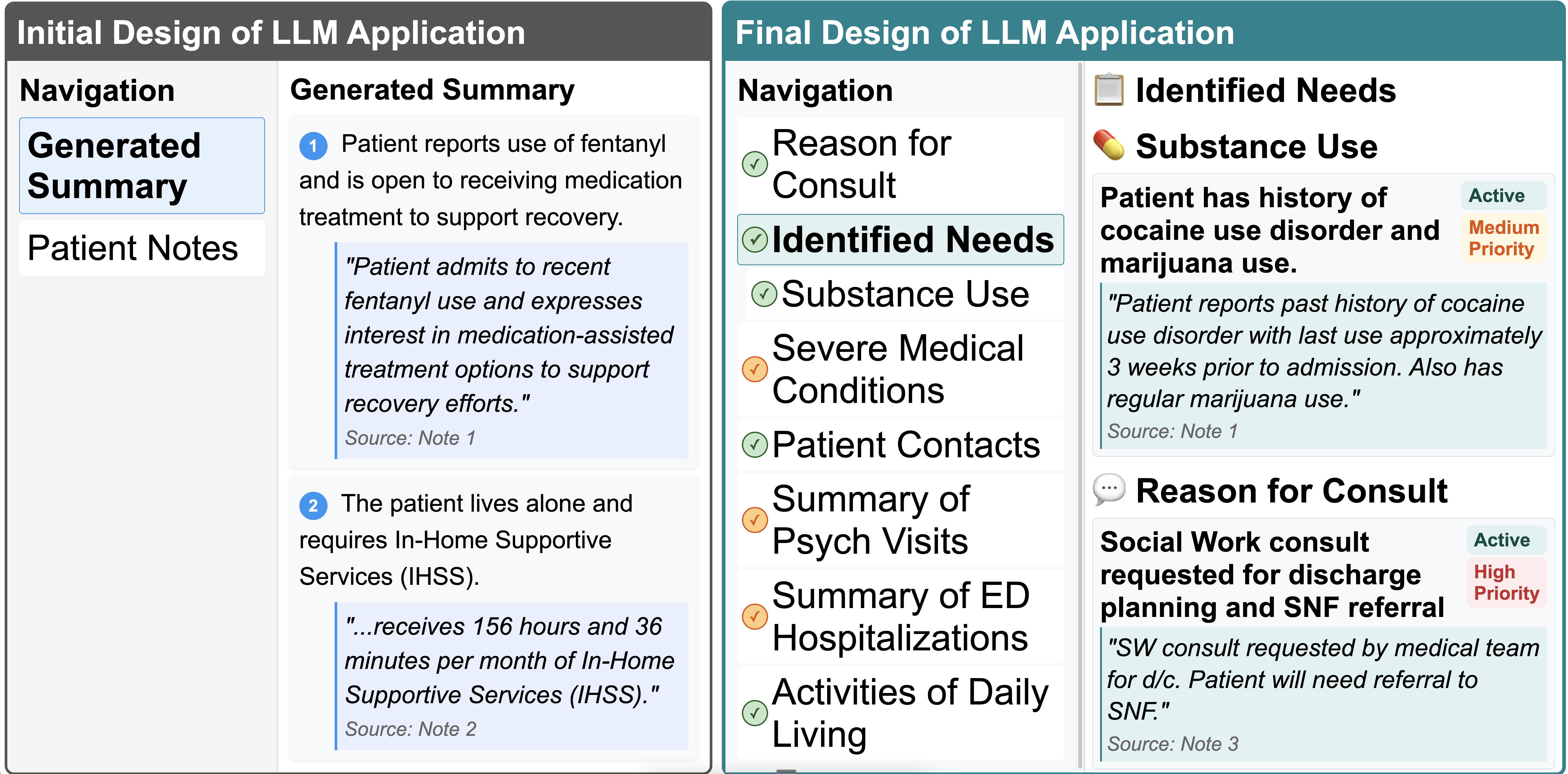}
    \caption{The initial design of the LLM pre-chart summarization application was to present a free-form summary (top). Given practical limitations in resource-constrained settings, the final design (bottom) presents a structured patient summary as \numattributes individual components, where components that underwent validation are highlighted by green checkmarks and attributes needing live verification are indicated by orange checkmarks.
    Supporting quotes from clinical notes are chosen to allow for easy verification during use of the application.
}
    \label{fig:app}
    \vspace{-0.2cm}
\end{figure}

\begin{tcolorbox}[boxsep=0pt, left=3pt, right=3pt, top=2pt]
\textbf{Design strategy}: 
Deconstruct free-form summaries into a set of attributes to be presented in a semi-structured format within the LLM application. Modularity provides the structure to independently tune, optimize, and validate attributes. LLMs also tend to extract information for attributes more accurately than free-form summaries. The user interface can also be designed in a modular fashion.

\end{tcolorbox}
\textbf{Implementation:} 
Based on the list of SBDH screening areas from the Centers for Medicare \& Medicaid Services (CMS) \citep{cms} and the day-to-day operational needs of the hospital SWs, the data science team decomposed the summarization task into \numattributes attributes (see final list in Appendix).
Although the attribute definitions were initially vague (i.e., a summary attribute may simply be defined as ``housing instability''), this modularization provided the overall structure for efficient downstream intention formation and prompt creation.
Furthermore, the user interface could now also reflect this modularity, where not all elements necessarily had to be human-validated. Rather, certain elements could be labeled as ``human-validated'' while others may highlight a need for more detailed verification by the SW during live use by reviewing quotes from the original clinical note (Figure~\ref{fig:app}).

\subsection{Step 3: Tiered refinement of summary attributes}
\begin{tcolorbox}[boxsep=0pt, left=3pt, right=3pt, top=2pt]
\textbf{Design strategy}:
Individually refine the attribute definitions and prompts using an iterative, multi-tier cascaded approach.
To initialize the refinement process with sufficiently well-defined attributes, bootstrap the process by assembling and repurposing readily available resources.
Then, moving up the domain expertise ladder, elicit concrete, data-driven feedback from CFEs and domain experts to refine attribute definitions and prompts.
\end{tcolorbox}

\textbf{Implementation}:
To make the most efficient use of resources, the team created a three-tier cascade.
The first tier involves DSs constructing sufficiently good attribute definitions and prompts to initialize the refinement process with CFEs and SWs.
The second tier involves CFEs working with DSs to check and refine attribute definitions and prompts and address as many points of confusion as possible.
In the third tier, SWs provide targeted, ``gold-standard'' feedback.
While generally the progression was from tier 1 to 3, the process was often cyclic, bouncing between tiers 1 and 2 before reaching tier 3.

\ul{Tier 1: Bootstrap from accessible organizational artifacts.} 
The DSs found two data sources at the safety-net hospital that could be assembled in a piecemeal fashion to initialize attribute definitions and prompts that were mature enough for the iterative refinement process with limited feedback.
The first was \textit{onboarding materials used to train new SWs}, which contains terminology and resources specific to the local context.
Despite the fact that onboarding documents were scattered across the organization and no single document comprehensively covered all social needs, DSs could use these to teach an LLM to tune attribute definitions and prompts; in fact, the team found this to be more effective than hand-translating the onboarding document themselves.

The second artifact was \textit{past SW notes}.
Although these notes are summaries of patient meetings and are not pre-chart summaries, these notes still provide useful example language and jargon for how different SBDH domains are discussed.
Quotes from these notes can thus serve as useful in-context learning (ICL) examples.
To find candidate quotes to include, DSs used an LLM to scrape notes for potentially useful quotes.
CFEs could also later help select among these quotes, which ones would be most useful to include as ICL examples.

\ul{Tier 2: Gather concrete feedback by asking CFEs to extract attributes.}
Once the attribute definitions and prompts reached a minimum level of maturity, CFEs were able to provide concrete feedback.
They were asked to annotate a small set of notes following the same instructions as the LLM, which helped highlight points of confusion and disagreements between the CFEs, DSs, and the LLM.
The CFEs and DSs were then able to collaboratively improve the clarity and specificity for most attributes.
Nevertheless, there still remained unresolved points of confusion that needed input from SWs.

\ul{Tier 3: Consult SWs on high value cases.} 
For attributes where extraction instructions caused confusion or there was high annotator disagreement, we asked for explicit feedback from the SWs. This targeted approach reduced the number of attributes requiring their input and improved the speed of prompt refinement. We elicited their feedback by asking for (i) a review of the attribute definitions and prompts for completeness and alignment with their workflow and (ii) help annotating a small subset of notes that CFEs had trouble with.
This led to further refinement of the summary attributes.
First, annotations by the expert taught team members about the subtle distinctions, which led to further refinement of the attribute definitions and prompts.
Second, certain attributes had to be updated because onboarding materials contained outdated information.
Finally, the attribute of ``interpersonal violence" (IPV) was found to require professional judgment that CFEs and DSs did not possess, which meant that validation of this attribute would not be feasible.
As such, the team decided to broaden the attribute to encompass all ``safety concerns" (e.g., anywhere from mild to more severe safety concerns).
This could be much more accurately identified and validated by the entire team, thereby de-risking the LLM application.

\begin{table*}
    \centering
    \footnotesize
    \begin{tabular}{p{2cm}|p{3cm}|p{3.4cm}|p{8cm}}
        Prompt1 & Prompt2 & Prompt3 & Prompt4 \\
        \toprule
       Review the notes and identify any mention of the social need category of Mental Health. Extract actions including previous, current, and planned actions.
       &
        Review the notes and identify any mention of the social need category of Mental Health. Extract actions including previous, current, and planned actions.  \textcolor{orange}{When extracting actions, look for interventions such as IMD facilities, LSAT, MHRC, ADU, psychiatric holds/conservatorship}
        & Review the notes and identify any mention of the social need category of \textcolor{orange}{Mental Health - Depression, anxiety, suicidal thoughts, or psychiatric symptoms impacting functioning.} Extract actions including previous, current, and planned actions.  When extracting actions, look for interventions such as IMD facilities, LSAT, MHRC, ADU, psychiatric holds/conservatorship
        & 
        Review the notes and identify any mention of the social need category of Mental Health - Depression, anxiety, suicidal thoughts, or psychiatric symptoms impacting functioning. Extract actions including previous, current, and planned actions. When extracting actions, look for interventions such as IMD facilities, LSAT, MHRC, ADU, and psychiatric holds/conservatorship.        
        \textcolor{orange}{Examples social needs category of ``mental health" (non exhaustive list):
          \begin{itemize}[topsep=0pt, partopsep=0pt, itemsep=0pt, parsep=0pt, leftmargin=*]
            \item Hallucinations (seeing or hearing things that are not there)
            \item Severe mood swings (e.g., manic or depressive episodes)
          \end{itemize}
        Examples where the social needs category of "mental health" should not be annotated to exist:
          \begin{itemize}[topsep=0pt, partopsep=0pt, itemsep=0pt, parsep=0pt, leftmargin=*]
            \item ``res was sent to ER for evaluation." 
            reason: mental health may not be the reason the resident was sent to the ER
            \item ``Resident indicated she is still feeling scared and relieved the resident was transferred to a different neighborhood on the unit" 
            reason: feeling scared does not qualify
       \end{itemize}}\\
       \bottomrule
    \end{tabular}
    \caption{Example of prompt tuning for a specific attribute using the tiered approach. Orange text highlights changes from the previous iteration. Each revision makes the extraction requirements more specific and clear. The transitions between prompts reflect the following: Prompt1 $\rightarrow$ Prompt2 incorporates interventions from SW onboarding documentation; Prompt2 $\rightarrow$ Prompt3 includes feedback from CFEs; Prompt3 $\rightarrow$ Prompt4 integrates fragmented examples drawn from previous SW notes}
    \label{tab:prompt_evolution}
    \vspace{-0.2cm}
\end{table*}

\subsection{Step 4: Tiered evaluation of LLM extractions}
\begin{tcolorbox}[boxsep=0pt, left=3pt, right=3pt, top=2pt]
\textbf{Design strategy}:
To track clarity of intermediate attribute prompts, use low-cost validation methods like LLM-as-a-judge.
To validate the extractions for the final attributes with high-cost human annotations, carefully plan and power the study using methods from experimental design.

\end{tcolorbox}

\textbf{Implementation}:
As the SWs had no bandwidth to assist with validation of the LLM extractions, there were only two tiers from the validation step.
Tier 1 involves DSs relying on LLM-as-a-judge as an approximate guide.
Tier 2 is a carefully designed human validation study with both CFEs and DSs annotating notes.

\ul{Tier 1: Evaluate clarity of intermediate prompts using LLM-as-a-judge.}
It is helpful to measure progress across each revision of the attribute definitions and prompts.
Because human annotations at each iteration are too expensive, we use an LLM-as-a-judge pipeline \cite{zheng2023judging}, in which an LLM was asked to judge the LLM application's extraction given the original patient note.
Generally, we expect attribute definitions and prompts that are more vague to have more disagreements between the LLM application and the LLM-as-a-judge.
So, an LLM-as-a-judge pipeline can approximately assess a prompt's clarity, with the benefit that the LLM-as-a-judge pipelines can be run at scale and minimal cost \citep{gu2024survey, zheng2023judging}.
We refer the reader to the Appendix for an example LLM judge prompt.

\ul{Tier 2: Adaptive sampling for efficient human evaluation.}
For the human annotation study, we leveraged well-established strategies from experimental design to minimize human annotation costs while maintaining statistical rigor.
First, we conducted a power analysis to determine the minimum sample size needed to quantify the LLM's performance with statistical confidence.
Our primary goal was to establish that the LLM's sensitivity was above predefined minimum threshold, $\tau$, which translates to testing the null hypothesis $H_0: \text{sensitivity} < \tau$. Using standard sample size calculations, one can then determine the minimum number of notes that needed to be annotated.

Second, we used adaptive sampling to ensure that all attributes would be adequately annotated, even the ones that are rarer.
However, classical approaches to adaptive sampling could not be directly applied because there were no structured data attributes to use for calculating sample weights.
As such, an LLM first extracted a probability $\hat{p}_{ij}$ of attribute $j$ appearing within note $i$ (a value between 0 and 1 means the LLM is unsure).
Then for attribute $j$, the probability of selecting note $i$ for inclusion is $\max\left(\rho_{\min}, \hat{p}_{ij}\right)$, where $\rho_{\min}$ (say 0.05) is a minimum selection probability.

\section{Experimental Results}
We present results from a tiered evaluation of the LLM application.
Experiments were run using a PHI compliant version of GPT 4o (OpenAI, 2024) on clinical notes for patients admitted at \iftoggle{submission}{\iftoggle{aaai}{\hospital{anonymous}}{\hospital{public}}}{\hospital{public}} between January 2023 to January 2024.
Patient summaries were generated using the following note types: Consults (Psychiatry, Social Work, Behavioral Health, Care Management, Complex Care Management), ED Provider Notes, Discharge Summaries, Progress Notes, and History \& Physical.
Summaries were based on notes from the two most recent encounters, with a maximum of 10 notes per note type.

\subsection{LLM-as-a-judge Results}
We first used LLM-as-a-judge to (approximately) evaluate attribute clarity across each iteration of prompt refinement.
To evaluate the LLM application across patients with varying levels of medical and social needs, we sampled 99 patients stratified by their note volume, resulting in a total of 1,216 notes.
We tested four prompt refinements: the initial attribute-level prompt written by the DSs (\promptone),  the prompt after incorporating onboarding materials for SWs (\prompttwo), then after receiving feedback from CFEs (\promptthree), and the final prompt that incorporated fragments from previous social work notes as ICL examples (\promptfour).

Table~\ref{tab:prompt_evolution} shows how tiered tuning of attribute definitions and prompts incrementally increased the concordance between the LLM application and LLM-as-a-judge.\footnote{Table~\ref{tab:prompt_evolution} shows 13 of the \numattributes attributes because two attributes were added at the end of the prompt tuning process but before human validation.} Despite limited access to domain experts, this tiered prompt tuning approach demonstrated a statistically significant improvement in almost every attribute, with an 18\% overall improvement in concordance with the LLM-judge (confidence intervals and p-values are shown in the Appendix). Several attributes increased in concordance substantially---by over 25\%---from initial to final prompt versions, including Outpatient Therapy and Reason for Consult. These attributes improved the most because their definitions require more specialized terminology and knowledge that had to be elicited from domain experts iteratively.
Finally, whereas the initial prompt exhibited high variability in LLM-judge concordance across the different attributes, the final prompt exhibits \textit{much} lower variability and thus better quality control.

\begin{table}[h]
\centering
\resizebox{\linewidth}{!}{\begin{tabular}{p{3.2cm}|>{\centering\arraybackslash}p{1.1cm}|>{\centering\arraybackslash}p{1.1cm}|>{\centering\arraybackslash}p{1.1cm}|>{\arraybackslash}p{1.1cm}}
 & \promptone & \prompttwo & \promptthree & \promptfour \\
\midrule
Alcohol Use & 0.68 & 0.88 & 0.93 & 0.94* \\
Medical Equipment & 0.91 & 0.97 & 0.94 & 0.99* \\
Food Insecurity & 1.00 & 0.98 & 0.98 & 0.99 \\
Housing & 0.75 & 0.77 & 0.94 & 0.93*\\
Immigration & 0.94 & 0.97 & 0.99 & 1.00* \\
Mental Health & 0.80 & 0.88 & 0.93 & 0.95*\\
Patient Contacts & 0.78 & 0.85 & 0.98 & 0.99*\\
Safety & 0.85 & 0.92 & 0.84 & 0.91 \\
Reason for Consult & 0.54 & 0.86 & 0.88 & 0.92*\\
Substance Use & 0.73 & 0.89 & 0.90 & 0.93*\\
Opiod Use & 0.82 & 0.76 & 0.87 & 0.88 \\
 Outpatient Therapy & 0.51 & 0.72 & 0.65 & 0.90*\\
Tobacco Use & 0.70 & 0.93 & 0.96 & 0.96*\\
\textbf{Overall/Average} & \textbf{0.77} & \textbf{0.88} & \textbf{0.91} & \textbf{0.95}*\\
\bottomrule
\end{tabular}
 }
\caption{Concordance between the LLM application and the LLM-as-a-judge across the patient summary attributes. Results are shown for four prompts that were iteratively refined using tiered attribute specification and prompt tuning approaches (ordered left to right). Here ``*'' denotes a statistically significant increase ($p < 0.05$) from \promptone{} to \promptfour{}.}
\label{tab:prompt_results}
\end{table}

\subsection{Human Validation Results}
After the prompt refinement process, we compared model extractions with human annotations for the final prompt (\promptfour). We asked four CFEs and DSs to annotate notes. Prior to annotation, the four annotators underwent training through three iterative sessions designed to address any ambiguities in attribute definitions. Between sessions, we assessed inter-annotator agreement to identify areas requiring clarification and engaged SWs to annotate complex social needs cases for additional annotator training. The final human annotation study involved 85 notes, selected through stratified random sampling (in expectation, each attribute should appear at least five times in the dataset). To assess inter-annotator reliability, we assigned 10\% of notes to multiple annotators.

Table~\ref{tab:annotator_results} presents three measures: agreement between annotators (\texttt{inter-annotator}), agreement between the LLM application and human annotators (\texttt{LLM app vs annotator}), and agreement between the human annotators and LLM-as-a-judge (\texttt{LLM judge vs annotator}).
We include a chance-corrected version of the agreement measure, Gwet’s Agreement Coefficient 1 \citep{Philip2025-sq}, in the Appendix.

The LLM application was highly accurate with an overall agreement rate with human annotators of 0.89, which is the same as the overall agreement rate between annotators.
This illustrates the potential of the LLM application to significantly improve the SW workflow. 
Furthermore, the overall agreement between the LLM-as-a-judge and annotators was even higher, achieving 0.92.
This demonstrates that LLM judges are generally helpful in measuring progress in prompt clarity.

If we further break down the results by attribute, we found that the agreement rate between LLM extractions and annotators exceeded 0.8 for nearly all attributes except for two: ``Patient Contacts'' and ``Reason for Consult."
These two attributes had notably lower agreement rates, even though the LLM-as-a-judge had scored their LLM extractions highly.
This shows that while high LLM-judge scores generally imply that a prompt is sufficiently clear and accurate, it is not always the case.
Indeed, the LLM-as-a-judge had some of the lowest agreement rates with the human annotator for these two attributes (0.88 and 0.79, respectively).
Thus, LLM-as-a-judge can be useful as a general guide but should not be treated as gold-standard. \textit{Human feedback and validation are still critical to have comprehensive quality control}.
Next steps for this LLM application are to obtain additional feedback on how to refine these two attributes and to test out the usability of the tool in real-world settings.
\begin{table}
\centering
\footnotesize
\resizebox{\linewidth}{!}{\begin{tabular}{p{3.5cm}|>{\centering\arraybackslash}p{1cm}|>{\centering\arraybackslash}p{1cm}|>{\centering\arraybackslash}p{1cm}}
& \spheading{inter-annotator}\ & \spheading{LLM app vs annotator} & \spheading{LLM judge vs annotator} \\
\midrule
Alcohol Use & 0.94 & 0.98 & 0.88 \\
Medical Equipment & 1.00 & 0.93 & 0.94 \\
Food Insecurity & 1.00 & 0.95 & 0.98 \\
Housing & 0.88 & 0.85 & 0.93 \\
Immigration & 1.00 & 0.98 & 1.00 \\
Mental Health & 0.63 & 0.81 & 0.88 \\
Patient Contacts & 0.88 & 0.72 & 0.88 \\
Safety & 0.81 & 0.86 & 0.92 \\
Reason for Consult & 0.81 & 0.65 & 0.79 \\
Substance Use & 0.94 & 0.88 & 0.88 \\
Opiod Use & 0.93 & 0.98 & 0.99 \\
Outpatient Therapy & 0.88 & 0.88 & 0.86 \\
Tobacco Use & 0.88 & 0.95 & 0.91 \\
Activities of Daily Living & 0.88 & 0.95 & 0.94 \\
Admission Reason & 0.88 & 0.84 & 0.85 \\
\textbf{Overall} & \textbf{0.89} & \textbf{0.89} & \textbf{0.92} \\
\bottomrule
\end{tabular} }
\caption{Agreement rates when extracting social need attributes between annotators, between the LLM application and annotators, and between annotators and the LLM-as-a-judge.}
\label{tab:annotator_results}
\vspace{-0.5cm}
\end{table}

\section{Discussion}

Co-design of LLM applications for resource-constrained settings introduces novel challenges.
There is often a wide gap between the AI developer and the domain expert's initial intentions to build an application and the final creation of such an application.
This gap involves numerous challenges, including iteratively clarifying the team's intentions and what exact information should be extracted, refining the corresponding prompts, and validating the final results.
The primary goal of this work is to highlight the need for new co-design strategies, as existing methods are too intensive for use in low-resource settings.
This manuscript presented a new multi-tier cascading framework that facilitates efficient, collaborative LLM application development.
Validation results demonstrate that this iterative methodology improved self-consistency across prompt iterations and high agreement rates with human annotators.
While this multi-tier framework was only demonstrated in this specific use case, we anticipate this framework, or at least its basic design strategies, could be useful in other under-resourced settings as well.

This study was conducted with IRB approval.
\iftoggle{submission}{
  \iftoggle{aaai}{
    Code is available at \repourl{anonymous}
  }{
    Code is available at \repourl{public}
  }
}{
  Code is available at \repourl{public}
}

\iftoggle{submission}{
  \iftoggle{aaai}{
  }{
    \section{Acknowledgements}
The PROSPECT lab (JF, LZ, AK, PV, JM) thanks Zuckerberg Priscilla Chan quality improvement fund via the San Francisco General Foundation for funding this project. We want to thank the entire \hospital{} Social Work Team for their invaluable feedback.
  }
}{
 \section{Acknowledgements}
The PROSPECT lab (JF, LZ, AK, PV, JM) thanks Zuckerberg Priscilla Chan quality improvement fund via the San Francisco General Foundation for funding this project. We want to thank the entire \hospital{} Social Work Team for their invaluable feedback.
}

\bibliographystyle{unsrt}
\bibliography{references}

\iftoggle{submission}{
\iftoggle{aaai}{
\section{Reproducibility Checklist}
This paper:

\begin{itemize}
    \item \textit{Includes a conceptual outline and/or pseudocode description of AI methods introduced} 
Yes - See section Experimental Results

    \item \textit{Clearly delineates statements that are opinions, hypothesis, and speculation from objective facts and results}
Yes - objective results are found in Experimental Results. Statements that are facts can be found with corresponding citations

    \item \textit{Provides well marked pedagogical references for less-familiare readers to gain background necessary to replicate the paper}
    Yes - While the dataset used in the manuscript is private, details of our co-design steps can be found in Resource-efficient LLM Co-Design and code can be found at \iftoggle{submission}{\iftoggle{aaai}{\repourl{anonymous}}{\repourl{public}}}{\repourl{public}} with details of experimental details found in the Experimental Results and Appendix

    \item \textit{Does this paper make theoretical contributions?}
No

    \item \textit{Does this paper rely on one or more datasets?}
    Yes - the dataset is described in the Experimental Results 

    \item \textit{If yes, please complete the list below.}

    \item \textit{A motivation is given for why the experiments are conducted on the selected datasets}
Yes - the problem we are addressing can be found in Problem Definition and the motivation for the experiments can be found in Resource-efficient LLM Co-Design

    \item \textit{All novel datasets introduced in this paper are included in a data appendix} 
No - the datasets used were a private dataset from \iftoggle{submission}{\iftoggle{aaai}{\hospital{anonymous}}{\hospital{public}}}{\hospital{public}} and the study was conducted with IRB approval 

    \item \textit{All novel datasets introduced in this paper will be made publicly available upon publication of the paper with a license that allows free usage for research purposes.}
No - the datasets used were a private dataset from \iftoggle{submission}{\iftoggle{aaai}{\hospital{anonymous}}{\hospital{public}}}{\hospital{public}} and were conducted with IRB approval 

    \item  \textit{All datasets drawn from the existing literature (potentially including authors’ own previously published work) are accompanied by appropriate citations.}
No - only a private dataset was used for this publication

    \item \textit{All datasets that are not publicly available are described in detail, with explanation why publicly available alternatives are not scientifically satisficing.}
    Yes - details of the dataset are found in Experimental Results and the problem motivation for using this dataset can be found in the Introduction and Problem Definition

    \item \textit{Does this paper include computational experiments?} Yes - in experimental results

    \item \textit{This paper states the number and range of values tried per (hyper-) parameter during development of the paper, along with the criterion used for selecting the final parameter setting.}

    Yes - details can be found in the Appendix Section Additional Experimental Details

    \item \textit{Any code required for pre-processing data is included in the appendix}
    Yes - details can be found in here \iftoggle{submission}{\iftoggle{aaai}{\repourl{anonymous}}{\repourl{public}}}{\repourl{public}}

    \item \textit{All source code required for conducting and analyzing the experiments is included in a code appendix} 
    Yes - details can be found in here \iftoggle{submission}{\iftoggle{aaai}{\repourl{anonymous}}{\repourl{public}}}{\repourl{public}}

    \item \textit{All source code required for conducting and analyzing the experiments will be made publicly available upon publication of the paper with a license that allows free usage for research purposes}
Yes, however the dataset will not be made public

    \item \textit{All source code implementing new methods have comments detailing the implementation, with references to the paper where each step comes from}
Yes code includes documentation detailing implementation and conducting experiments

    \item \textit{If an algorithm depends on randomness, then the method used for setting seeds is described in a way sufficient to allow replication of results}
Yes details can be found \iftoggle{submission}{\iftoggle{aaai}{\repourl{anonymous}}{\repourl{public}}}{\repourl{public}}

    \item \textit{This paper specifies the computing infrastructure used for running experiments (hardware and software), including GPU/CPU models; amount of memory; operating system; names and versions of relevant software libraries and frameworks}
Yes - details can be found in the Appendix Section Additional Experimental Details. Relevant libraries and frameworks can be found at \iftoggle{submission}{\iftoggle{aaai}{\repourl{anonymous}}{\repourl{public}}}{\repourl{public}}

    \item \textit{This paper formally describes evaluation metrics used and explains the motivation for choosing these metrics.}
Yes - evaluation metrics can be found in Experimental Results and additional evaluation metrics can be found in the Appendix Section Additional Experimental Details

    \item \textit{This paper states the number of algorithm runs used to compute each reported result.}
    Yes - details can be found in the Appendix Section Additional Experimental Details 

    \item \textit{Analysis of experiments goes beyond single-dimensional summaries of performance (e.g., average; median) to include measures of variation, confidence, or other distributional information}
    Yes - Table~\ref{tab:prompt_results} highlights statistically significant with confidence intervals and p values found in the Appendix Section Additional Experimental Details
  
    \item The significance of any improvement or decrease in performance is judged using appropriate statistical tests (e.g., Wilcoxon signed-rank). Yes - Table~\ref{tab:prompt_results} highlights statistically significant with confidence intervals and p values found in the Appendix. Table~\ref{tab:annotator_results} details additional metrics in the Appendix Section Additional Experimental Details

    \item \textit{This paper lists all final (hyper-)parameters used for each model/algorithm in the paper's experiments}
    Yes - details can be found in the Appendix Section Additional Experimental Details

\end{itemize}
}{}
}{}

\appendix
\setcounter{table}{0}
\renewcommand{\thetable}{A\arabic{table}}

\title{
When the Domain Expert Has No Time and the LLM Developer Has No Clinical Expertise: Real-World Lessons from LLM Co-Design in a Safety-Net Hospital\\
\textit{Supplementary Material}
}

\maketitle

\onecolumn

% \appendix \label{appendix}

\clearpage
\appendix
\begin{center}
\Huge\textbf{Supplementary Material}\\[3cm]
\end{center}

\section{Structured Summary Attributes}

{\footnotesize
\begin{longtable}{p{4cm}|p{10cm}|p{2cm}}

\hline
\textbf{Category} & \textbf{Description} & \textbf{Validated}\\
\endfirsthead

\midrule
Reason for Admission & The primary reason for the patient's admission for the current hospital encounter & X \\
\midrule
Reason for Consult & For social worker consult note types only, identify reason for the social work consult. If the consult is specified, identify if it should be classified as high priority by determining if it meets any of the following categories: Homeless and discharge ready, Jane/John Doe, Suffering from Abuse/neglect/self-neglect/safety,  Need Family Located, Death and Dying Case, Major Life Changing Event & X \\
\midrule
Patient's Contacts & Extraction of all contacts in a patient's support system where there is name and a phone number or email address. And any mentions of missing or minimal contacts or patient being socially isolated. Examples include mentions of: widowed, lives alone with no explict mention of family support, no family involved, no next of kin, family not reachable, no emergency contact list, expresses feelings of being alone or unsupported
& X\\
\midrule
Outpatient Therapies &  Identify if the patient has health workers that come to the patient’s home or are in a Skilled Nursing Facility (SNF) or a referral has been made to a SNF (not a recommendation). Examples include: Home health such as PT/OT/SLP/RN/ SW/ home health aide that come to the patient’s home & X \\
\midrule
    Opiod Use &  
    \begin{itemize}
        \item Date of last use (if available)
        \item Identify if the patient has a problematic pattern of opioid use (including fentanyl or heroin) leading to clinically significant impairment distress or the note mentions the patient has “regular opioid use over years” ir the note explicitly mentions the patient has OUD.(yes, no, unknown)
        \item Identify if the patient has  a problematic pattern of amphetamine-type substances, cocaine, methamphetamine, or other stimulant use leading to clinically significant impairment distress or the note mentions the patient has “regular meth use over years” or the note explicitly mentions the patient has StUD (yes, no, unknown)
        \item Polysubstance Use Disorder (PSUD) - The note explicitly mentions the patient has or has had PSUD
        \item Medication for Opioid Use Disorder (MOUD) - The note mentions the patient is on methadone or buprenorphine (i.e. suboxone, Brixadi, Sublocade). Abbreviations: bup (yes, no, unknown)
    \end{itemize} & X \\
    \midrule
Safety & Extraction and summary of the patient's social needs and previous, current, and planned interventions for the following social needs: 
\begin{itemize}
    \item Safety - Any form of abuse, domestic violence (DV), assault, sexual assault, being victim/perpetrator of violence, interpersonal violence, recurrent falls or unexplained injuries, or self neglect/ caregiver (manifest as an inability or unwillingness to perform essential self-care tasks, maintain personal hygiene, manage medical conditions)  
\end{itemize} & X\\
\midrule
Tobacco Use & Extraction and summary of the patient's social needs and previous, current, and planned interventions for the following social needs: 
\begin{itemize}
    \item Tobacco Use - Any  use of cigarettes, vaping, chewing tobacco, or other nicotine product 
\end{itemize} & X\\
\midrule
Housing & Extraction and summary of the patient's social needs and previous, current, and planned interventions for the following social needs: 
\begin{itemize}
    \item Housing - Homelessness, housing instability, unsafe living conditions, or inability to afford housing
\end{itemize} & X\\
\midrule
Immigration & Extraction and summary of the patient's social needs and previous, current, and planned interventions for the following social needs: 
\begin{itemize}
    \item Immigration - Documentation concerns, deportation fears, language barriers, or immigration-related discrimination.
\end{itemize} & X\\
\midrule
Food Insecurity & Extraction and summary of the patient's social needs and previous, current, and planned interventions for the following social needs: 
\begin{itemize}
    \item Food Insecurity - Inability to access adequate food due to financial constraints or running out of food
\end{itemize} & X\\
\midrule
Substance Use & Extraction and summary of the patient's social needs and previous, current, and planned interventions for the following social needs: 
\begin{itemize}
    \item Substance Use - Use of illegal drugs, prescription drug misuse, or problematic controlled substance use (does not include alcohol)
\end{itemize} & X\\
\midrule
Alcohol Use & Extraction and summary of the patient's social needs and previous, current, and planned interventions for the following social needs: 
\begin{itemize}
    \item Alcohol Use - Excessive drinking, binge drinking, or alcohol-related problems affecting daily life 
\end{itemize} & X\\
\midrule
Mental Health & Extraction and summary of the patient's social needs and previous, current, and planned interventions for the following social needs: 
\begin{itemize}
    \item Mental Health - Depression, anxiety, suicidal thoughts, or psychiatric symptoms impacting functioning. 
\end{itemize} & X\\
\midrule
Medical Equipment & Identify each piece of durable medical equipment the patient currently uses or requires. Examples include:
	mobility devices, wheelchairs, walkers, canes, crutches, Electric hospital bed, transfer lift with slings, Bedside commode, shower chair, tub bench, raised toilet seat with arms, grab bars, Home oxygen concentrator, portable Oxygen cylinders, nebulizer, CPAP or BiPAP unit, portable suction machine, Negative-pressure wound-vac pump, ambulatory,elastomeric infusion pump, IV pole
 	(e.g., walker, wheelchair) & X\\
\midrule
    Activities of Daily
Living & Identify any inability to perform ADLs (Activities of Daily Living) which refer to basic self care tasks bathing, dressing, toileting, grooming, eating. & X \\
    \midrule
Housing Situation &   
\begin{itemize}
    \item Rent or Own - Individual has stable, independent housing with legal tenancy rights through rental agreement or property ownership. SROs  (Single Room Occupancy - shared bathrooms and sometimes shared kitchens) if they do not have onsite
     \item Rent Unit with Support Services Onsite - Subsidized or supportive housing with integrated services (case management, mental health, etc.) available within the building
     \item Residential Treatment / Jail - will exit homeless 
     \item Living Outside (Street / Vehicle / Makeshift) - Street - Staying outdoors (on the street, sidewalk, doorway, park, freeway underpass). Vehicle - Staying in a car, van, bus, truck , RV, or similar vehicle. Makeshift - Staying in an enclosure or structure not authorized for habitation by building or housing codes (abandoned buildings)
     \item Temporarily with Friend / Family - Unstable housing arrangement staying with others.
     \item Stabilization Room / Hotel -  Living in a hotel room and does not have tenancy rights 
     \item Shelter / Navigation Center - taying in a supervised public or private facility that provides temporary living conditions (e.g. homeless shelter or mission)
     \item Permanent Supportive Housing - Residential Care Facility/Board and Care Facility. SROs if they do not have onsite services.
     \item Residential Treatment / Jail - will exit housed 
     \item Unknown
   \end{itemize} & \\
   \midrule
   Inpatient Therapies & Identify explicit mentions of the patient in the following inpatient therapies or have consults (not a recommendation). (PT, OT, SLP often fall under "Rehab" or "Speech"):
   \begin{itemize}
       \item Rehab
       \item Speech
       \item Nutrition
       \item Language
   \end{itemize}
   For each therapy mentioned:
   a. Confirm if there's a consult for the therapy (yes, no, or unknown)
   b. Provide a brief summary of the patient's therapy progress& \\
   \midrule
Severe Medical Conditions & Extraction and summary of medical conditions that require complex discharge planning, complicate transition of care, or may necessitate specialized care after discharge & \\
    \midrule
    ED Visits and Psychiatric Visits & Summary of any recurring patterns for both ED visits and Psychiatric Visits \\
    \bottomrule
    \caption{A full list of structured attributes extracted from patient notes to create LLM generated summaries for clinical social workers. Validated attributes were evaluated against human annotations to measure the accuracy of the LLM extractions.}
\end{longtable}
}

\section{Experimental Details}

\subsection{Implmentation Details}
All experiments were conducted using a PHI-compliant version of GPT-4o (OpenAI, 2024-08-06) accessed via API. Model hyperparameters were standardized across all experiments with a fixed seed value of 0 and temperature setting of 0, ensuring deterministic outputs with a single LLM extraction performed per patient, per attribute, per clinical note.

The experiments were run on r7i.2xlarge AWS EC2 machine with 
an Ubuntu operating system. The instance has 8 CPUs and 64 GBs of memory (no GPUs).

\subsection{Example Prompts}

\subsection{\promptone{}}

You are a clinical social worker with 10 years of hospital experience in San Francisco. Your task is to analyze a series of social worker notes to generate a structured JSON summary of the patient for a clinical social worker. For all extracted information in the summary, include a citation with the note ID and an exact quote from the note that supports your evidence.

Instructions:
\begin{enumerate}
\item Review the notes and identify any mention of these specific social needs:
\begin{itemize}
    \item Interpersonal violence
   \item Tobacco use
   \item Housing
   \item Immigration
   \item Food insecurity
   \item Substance use
   \item Alcohol use
   \item Mental health
\end{itemize}

\item  For each identified social need, provide:
   a. A brief description of the social need (specific details about the patient's situation)
   b. A classification of the need's status (must be exactly one of: ``resolved", ``ongoing", ``new") 
   c. A priority level (must be exactly one of: ``high", ``medium", ``low")

\item Extract actions corresponding to each identified need. Categorize actions as:
   a. Previous actions (completed):
      - Description of what was done
      - Start date (if available)
      - Outcome
   b. Current actions (in progress):
      - Description of what is being done
      - Start date (if available)
   c. Planned actions:
      - Description of what needs to be done

\item Identify the patient's current and past housing situation as one of these exact categories:
   - Rent or Own
   - Rent Unit with Support Services Onsite
   - Residential Treatment / Jail - will exit homeless
   - Living Outside (Street / Vehicle / Makeshift)
   - Temporarily with Friend / Family
   - Stabilization Room / Hotel / SRO
   - Shelter / Navigation Center
   - Permanent Supportive Housing
   - Residential Treatment / Jail - will exit housed
   - Unknown
  a. Provide a brief description of this patient's housing situation

\item Identify the reason for the social work consult and if it is a high, medium, or low priority consult
   High-priority consults include:
   - Homeless and discharge ready
   - Jane/John Doe
   - Suffering from Abuse/neglect/self-neglect/safety
   - Need family located

\item Identify the reason for the admission

\item Identify any diagnoses for the current encounter

\item Identify mentions of these therapies or consults:
   - Rehab
   - Speech
   - Nutrition
   - Language
   For each therapy mentioned:
   a. Confirm if there's a consult for the therapy (yes, no, or unknown)
   b. Provide a brief summary of the patient's therapy progress

    \item Identify the patient's contacts:
   For each contact, provide name, phone number, and relation (if known)

    \item Provide a brief description of the patient's support at home

    \item If the patient has a history of substance use, identify the following:
    a. A brief temporal summary describing the patient's substance use 
    b. Date of last use (if available)
    c. Whether patient is currently on methadone (yes, no, unknown)

    \item Identify any severe medical conditions that could affect discharge, including diagnosis and date if available

    \item Provide a brief description of any recurring patterns for ED visits

    \item Provide a brief description of any recurring patterns for psychiatric hospitalizations

    \item Identify any durable medical equipment (DME) the patient uses (e.g., walker, wheelchair)
    a. A brief description describing the reason for using this durable medical equipment

\end{enumerate}

\subsection{\prompttwo{}}

You are a highly experienced Clinical Social Worker. You have 10 years of specialized experience in inpatient hospital social work, adept at managing:
-   Discharge Home Arrangements (e.g., community case management, DME, home care, meal services)
-   Facility Transfers (e.g.,  psychiatric care, substance disorder treatment)
-   Transportation Logistics (e.g., wheelchair van, BLS ambulance, paratransit)

Your task is to analyze patient notes and extract key patient information into structured JSON format. Every extracted piece of information must include a citation with note ID and exact supporting quote.

Instructions:
\begin{enumerate}

\item Review the notes and identify any mention of these specific social needs:
\begin{itemize}
    \item Interpersonal violence
   \item Tobacco use
   \item Housing
   \item Immigration
   \item Food insecurity
   \item Substance use
   \item Alcohol use
   \item Mental health
\end{itemize}

\item  For each identified social need, provide:
   a. A brief description of the social need (specific details about the patient's situation) and include any temporal changes. 
   b. A classification of the need's status (must be exactly one of: ``resolved",``ongoing", ``new") 
   c. A priority level (must be exactly one of: ``high", ``medium", ``low")

\item Extract actions corresponding to each identified need. Categorize actions as:
   a. Previous actions (completed):
      - Description of what was done
      - Start date (if available)
      - Outcome
   b. Current actions (in progress):
      - Description of what is being done
      - Start date (if available)
   c. Planned actions:
      - Description of what needs to be done

  When extracting actions, look for interventions such as:
  - Mental Health: LSAT, MHRC, ADU, psychiatric holds/conservatorship
  - Substance Use: residential/outpatient SUD programs
  - Housing/Discharge: Home health (PT/OT/SLP/RN), IHSS caregivers, shelters, navigation centers, Medical Respite, Coordinated Entry (CE) assessments, SNF referrals, Board and Care/ALF/RCFE, ARU/ARF, LTACH

\item Identify the patient's current and past housing situation as one of these exact categories:
\begin{itemize}
    \item Rent or Own
    \item Rent Unit with Support Services Onsite
    \item Residential Treatment / Jail - will exit homeless
    \item Living Outside (Street / Vehicle / Makeshift)
    \item Temporarily with Friend / Family
    \item Stabilization Room / Hotel / SRO
    \item Shelter / Navigation Center
    \item Permanent Supportive Housing
    \item Residential Treatment / Jail - will exit housed
    \item Unknown
\end{itemize}

\begin{enumerate}[label=\alph*.]
    \item Provide a brief description of this patient's housing situation. Include information such as home infusions or caregivers. If the patient is in coordinate entry include the following information:
    \begin{itemize}
        \item Accessing Coordinated Entry
        \begin{itemize}
            \item Problem Solving Status
            \item Housing Referral Status
            \item Housing Navigation / Move-in
            \item Housed (through HRS or outside)
        \end{itemize}
        \item Include information about Administrative Review, Reasonable Accommodation, Problem Solving funds, Housing Navigator outreach, or referral to PSH/RRH. Include information about start dates and outcomes
    \end{itemize}
\end{enumerate}

\item Identify if the social work consult is high priority. Answer yes if it is high priority otherwise answer no.

High-priority consults include:
\begin{itemize}
    \item Homeless and discharge ready: Patients who are medically cleared for discharge but have no stable housing to return to. Includes shelter residents, those living in places not meant for habitation, those who cannot return to their previous housing.
    
    \item Jane/John Doe: Patients whose identity cannot be verified or established. Includes unconscious patients without ID, those unable to communicate their identity, those providing unverifiable identifying information.
    
    \item Suffering from Abuse/neglect/self-neglect/safety
    \begin{itemize}
        \item Abuse: Evidence of physical, emotional, sexual, financial, or psychological harm by others
        \item Neglect: Inadequate care by caregivers resulting in harm or risk
        \item Self-Neglect: Patient's failure to meet their own basic needs, endangering health
        \item Safety: Conditions that compromise patient safety at home or upon discharge
    \end{itemize}
    
    \item Need family located: Cases requiring identification and contact with family members or next of kin for purposes including medical decision-making, discharge planning, or providing critical medical updates
    
    \item Death and Dying: Any mention that a patient has died, is actively dying or that clinicians/family are engaging in end-of-life planning (e.g., hospice enrollment, DNR discussions, comfort-focused care)
    
    \item Major Life Changing Event: A sudden event that permanently alters the patient's daily functioning or psychosocial status and requires new resources/long-term planning. Includes new permanent disability, life-altering diagnosis, loss of primary caregiver or housing explicitly described as permanent), or other irreversible changes (e.g., limb amputation, newly paralyzed)
\end{itemize}

\item  Identify the reason for the admission

\item  Identify any diagnoses for the current encounter

\item  Identify mentions of these therapies or consults:
   - Rehab
   - Speech
   - Nutrition
   - Language
   For each therapy mentioned (e.g., PT, OT, SLP often fall under ``Rehab" or ``Speech"):
   a. Confirm if there's a consult for the therapy (yes, no, or unknown)
   b. Provide a brief summary of the patient's therapy progress

\item  Identify the patient's contacts:
   - For each contact, provide name, phone number, and relation (if known)

\item  Provide a brief description of the patient's support at home

\item  If the patient has a history of substance use, identify the following:
    a. A brief temporal summary describing the patient's substance use 
    b. Date of last use (if available)
    c. Whether patient is currently on methadone (yes, no, unknown)

\item  Identify any severe medical conditions that could affect discharge, including diagnosis and date if available

\item  Provide a brief description of any recurring patterns for ED visits

\item  Provide a brief description of any recurring patterns for psychiatric visits. Include if the patient typically presents or is referred to as ``acute", ``acute diversion" (e.g., ADU), ``respite" 

\item Identify any durable medical equipment (DME) the patient uses. Examples include: mobility devices, wheelchairs, walkers, canes, crutches Electric hospital bed, Hoyer/transfer lift with slings, Bedside commode, shower chair/tub bench, raised toilet seat with arms, grab bars Home oxygen concentrator, portable Oxygen cylinders, nebulizer, CPAP/BiPAP unit, portable suction machine, Negative-pressure wound-vac pump, ambulatory/elastomeric infusion pump, IV pole
    a. A brief description describing the reason for using this durable medical equipment
\end{enumerate}

\subsection{\promptthree{}}
You are a clinical social worker with 10 years of hospital experience. Your task is to analyze clinical note(s) to generate a structured JSON summary of the patient for a clinical social worker. For all extracted information in the summary, include a citation with the note ID and an exact quote from the note that supports your evidence

\begin{itemize}
    \item Discharge Home Arrangements (e.g., community case management, DME, home care, meal services)
    \item Facility Transfers (e.g., SNF referrals, psychiatric care, substance disorder treatment)
\end{itemize}

Instructions:

\begin{enumerate}
    \item Review the notes and identify any mention of these specific social needs:
    \begin{itemize}
        \item Safety - Any form of abuse, domestic violence (DV), assault, sexual assault, being victim/perpetrator of violence, interpersonal violence, recurrent falls or unexplained injuries, or self neglect/ caregiver
        \item Tobacco use - Any use of cigarettes, vaping, chewing tobacco, or other nicotine product
        \item Housing - Homelessness, housing instability, unsafe living conditions, or inability to afford housing
        \item Immigration - Documentation concerns, deportation fears, language barriers, or immigration-related discrimination
        \item Food insecurity - Inability to access adequate food due to financial constraints or running out of food
        \item Substance use - Use of illegal drugs, prescription drug misuse, or problematic controlled substance use
        \item Alcohol use - Excessive drinking, binge drinking, or alcohol-related problems affecting daily life
        \item Mental health - Depression, anxiety, suicidal thoughts, or psychiatric symptoms impacting functioning
    \end{itemize}

    \item For each identified social need, provide:
    \begin{enumerate}[label=\alph*.]
        \item A brief description of the social need (specific details about the patient's situation)
        \item A classification of the need's status (must be exactly one of: ``resolved'' - Need was present but has been addressed and is no longer active, ``ongoing'' - Need continues to be present, ``new'' - Need has been newly identified in this encounter)
    \end{enumerate}

    \item Extract actions corresponding to each identified need. Categorize actions as:
    \begin{enumerate}[label=\alph*.]
        \item Previous actions - Interventions (these should be not recommendations) that were implemented in the past - referrals, connections, or consults previously made:
        \begin{itemize}
            \item Description of what was done
            \item Start date (if available)
            \item Outcome
        \end{itemize}
        \item Current actions (in progress) - Interventions (these should be not recommendations) - referrals, connections, or consults that are actively in progress. A referral was put in or the patient is actively using that service currently made or in progress:
        \begin{itemize}
            \item Description of what is being done
            \item Start date (if available)
        \end{itemize}
        \item Planned actions - Interventions (these should be not recommendations) that are scheduled or intended for the future - referrals, connections, or consults that will be made:
        \begin{itemize}
            \item Description of what needs to be done
        \end{itemize}
    \end{enumerate}

    When extracting actions, look for interventions such as (this is a nonexhaustive list):
    \begin{itemize}
        \item Mental Health: IMD facilities, LSAT, MHRC, ADU, psychiatric holds/conservatorship
        \item Substance Use: detox or withdrawal management referrals, residential/outpatient SUD programs, Addiction Care Team Patient Navigator
        \item Housing/Discharge: Home health (PT/OT/SLP/RN), shelters, navigation centers, Medical Respite, Coordinated Entry (CE) assessments, SNF referrals, Board \& Care/ALF/RCFE, ARU/ARF, LTACH
    \end{itemize}

    \item Identify the patient's current housing situation as one of these exact categories:
    \begin{itemize}
        \item Rent or Own - Individual has stable, independent housing with legal tenancy rights through rental agreement or property ownership. SROs (Single Room Occupancy - shared bathrooms and sometimes shared kitchens) if they do not have onsite
        \item Rent Unit with Support Services Onsite - Subsidized or supportive housing with integrated services (case management, mental health, etc.) available within the building
        \item Residential Treatment / Jail - will exit homeless - Individual currently in institutional setting who will have no housing upon discharge/release
        \item Living Outside (Street / Vehicle / Makeshift) - Street - Staying outdoors (on the street, sidewalk, doorway, park, freeway underpass). Vehicle - Staying in a car, van, bus, truck, RV, or similar vehicle. Makeshift - Staying in an enclosure or structure not authorized for habitation by building or housing codes (abandoned buildings)
        \item Temporarily with Friend / Family - Unstable housing arrangement staying with others
        \item Stabilization Room / Hotel - Living in a hotel room and does not have tenancy rights
        \item Shelter / Navigation Center - Staying in a supervised public or private facility that provides temporary living conditions (e.g. homeless shelter or mission)
        \item Permanent Supportive Housing - Residential Care Facility/Board and Care Facility. SROs if they do not have onsite services
        \item Residential Treatment / Jail - will exit housed - Individual in institutional setting who has secured housing arrangement upon discharge/release
        \item Unknown
    \end{itemize}
    \begin{enumerate}[label=\alph*.]
        \item Provide a brief description of this patient's housing situation
    \end{enumerate}

    \item For social worker consult note types only, Identify the reason for the social work consult. Do not extract this information if the note is not a social work consult note type

    \item Identify the reason for the admission

    \item Outpatient Therapy and Home Health Services - Identify if the patient has health workers that come to the patient's home or are in a Skilled Nursing Facility (SNF) or a referral has been made to a SNF. Examples include: Home health such as PT/OT/SLP/RN/ SW/ home health aide that come to the patient's home or IHSS caregivers (not a family member)

    \item Identify mentions of these inpatient therapies or consults (not a recommendation). (PT, OT, SLP often fall under ``Rehab'' or ``Speech''):
    \begin{itemize}
        \item Rehab
        \item Speech
        \item Nutrition
        \item Language
    \end{itemize}
    For each therapy mentioned:
    \begin{enumerate}[label=\alph*.]
        \item Confirm if there's a consult for the therapy (yes, no, or unknown)
        \item Provide a brief summary of the patient's therapy progress
    \end{enumerate}

    \item Identify any references to a patient's contacts. Ex: any mention of friends, family, case workers:
    \begin{itemize}
        \item For each contact, provide name, phone number, and relation (if known)
    \end{itemize}

    \item A brief description of missing or minimal contacts or patient being socially isolated. Examples include mentions of: widowed, lives alone, no family involved, no next of kin, family not reachable, no emergency contact list, expresses feelings of being alone or unsupported.
    
    Examples:
    \begin{itemize}
        \item ``The patient is widowed and lives alone.''
        \item ``No family involved; the patient has no next of kin.''
        \item ``Family members are not reachable.''
        \item ``No emergency contact list available.''
        \item ``The patient expresses feelings of being alone or unsupported.''
    \end{itemize}

    \item Identify any inability to perform ADLs (Activities of Daily Living) which refer to basic self care tasks bathing, dressing, toileting, grooming, eating.
    \begin{enumerate}[label=\alph*.]
        \item A brief description of the ADLs
    \end{enumerate}

    \item Identify any Inability to perform IADLs (Instrumental Activities of Daily Living) which refer to more complex self care tasks. managing medications, using transportation, cooking, handling finances, shopping, housekeeping, managing appointments
    \begin{enumerate}[label=\alph*.]
        \item A brief description of the IADLs
    \end{enumerate}

    \item If the patient has a history of substance use or is currently struggling with substance use, identify the following:
    \begin{enumerate}[label=\alph*.]
        \item A brief temporal summary describing the patient's substance use
        \item Date of last use (if available)
        \item Identify if the patient has a problematic pattern of opioid use (including fentanyl or heroin) leading to clinically significant impairment distress or the note mentions the patient has ``regular opioid use over years'' or the note explicitly mentions the patient has OUD (yes, no, unknown)
        \item Identify if the patient has a problematic pattern of amphetamine-type substances, cocaine, methamphetamine, or other stimulant use leading to clinically significant impairment distress or the note mentions the patient has ``regular meth use over years'' or the note explicitly mentions the patient has StUD (yes, no, unknown)
        \item Polysubstance Use Disorder (PSUD) - The note explicitly mentions the patient has or has had PSUD
        \item Medication for Opioid Use Disorder (MOUD) - The note mentions the patient is on methadone or buprenorphine (i.e. suboxone, Brixadi, Sublocade). Abbreviations: bup (yes, no, unknown)
    \end{enumerate}

    \item Identify each piece of durable medical equipment the patient currently uses or requires. Examples include:
    \begin{itemize}
        \item mobility devices, wheelchairs, walkers, canes, crutches
        \item Electric hospital bed, Hoyer/transfer lift with slings, Bedside commode, shower chair/tub bench, raised toilet seat with arms, grab bars
        \item Home oxygen concentrator, portable Oxygen cylinders, nebulizer, CPAP/BiPAP unit, portable suction machine
        \item Negative-pressure wound-vac pump, ambulatory/elastomeric infusion pump, IV pole
    \end{itemize}
    (e.g., walker, wheelchair)
    \begin{enumerate}[label=\alph*.]
        \item A brief description describing the reason for using this durable medical equipment
    \end{enumerate}

    \item Provide a brief description of any recurring patterns for ED visits

    \item Provide a brief description of any recurring patterns for psychiatric visits. Include if the patient typically presents or is referred to as ``acute'', ``acute diversion'' (e.g., ADU), ``respite'' 

    \item Provide a brief description of the patient's support at home

    \item Identify if this patient is high priority (yes, no, unknown) and provide your reasoning.
    
    High-priority patients include at least one of the following:
    \begin{itemize}
        \item Homeless and discharge ready - Patients who are medically cleared for discharge but have no stable housing to return to. Includes shelter residents, those living in places not meant for habitation, those who cannot return to their previous housing.
        \item Jane/John Doe - Patients whose identity cannot be verified or established. unconscious patients without ID, those unable to communicate their identity. those providing unverifiable identifying information.
        \item Suffering from Abuse/neglect/self-neglect/safety - Abuse: Evidence of physical, emotional, sexual, financial, or psychological harm by others, Neglect: Inadequate care by caregivers resulting in harm or risk, Self-Neglect: Patient's failure to meet their own basic needs, endangering health, Safety: Conditions that compromise patient safety at home or upon discharge
        \item Need family located - Cases requiring identification and contact with family members or next of kin for purposes including medical decision-making, discharge planning, or providing critical medical updates
        \item Death and Dying - Any mention that a patient has died, is actively dying or that clinicians/family are engaging in end-of-life planning (e.g., hospice enrollment, DNR discussions, comfort-focused care)
        \item Major Life Changing Event - A sudden event that permanently alters the patient's daily functioning or psychosocial status and requires new resources/long-term planning. Includes new permanent disability, life-altering diagnosis, loss of primary caregiver or housing explicitly described as permanent), or other irreversible changes (e.g., limb amputation, newly paralyzed)
    \end{itemize}
\end{enumerate}

\subsection{\promptfour{}}
\textit{Role.}  
You are a clinical social worker with 10 years of hospital experience.

\textit{Task.}  
Analyze clinical note(s) to generate a structured JSON summary of the patient for a clinical social worker. Do not make any leaps in logic; extract only the information clearly provided in the clinical notes. For every piece of extracted information in the summary,  
\begin{enumerate}
\item include a citation with the note ID and a verbatim excerpt from the note that supports your evidence
\end{enumerate}

If a quotation in the note falls under more than one category, include it in each relevant category. For instance, if the patient has a recent history of falls, it should be noted as both a current safety social need and a high priority case.

Area of Focus  
\begin{itemize}
\item Discharge Home Arrangements (e.g., community case management, DME, home care, meal services)
\item Facility Transfers (e.g., SNF referrals, psychiatric care, substance disorder treatment)
\end{itemize}

Instructions  
\begin{enumerate}
\item Review the notes and identify any mention of these specific social needs:
\begin{itemize}
\item Safety – Any indication of abuse, domestic conflict, assault, sexual assault, involvement in violence (as victim or perpetrator), repeated falls or unexplained injuries, or self-neglect (or neglect by a caregiver manifested as inability or unwillingness to perform essential self-care tasks including hygiene, medication management, etc.)

Examples where the social needs category of “safety” should be annotated:
\begin{itemize}
\item ...
\end{itemize}

Examples where the social needs category of “safety” should not be assigned:
\begin{itemize}
\item ...
\end{itemize}

\item Tobacco use – Any use of cigarettes, vaping devices, chewing tobacco, or other nicotine products
\item Housing – Instances of homelessness, housing instability, unsafe living conditions, or inability to afford stable housing
\item Immigration – Concerns regarding documentation, fears of deportation, language barriers, or discrimination related to immigration status
\item Food insecurity – Difficulty accessing sufficient food due to financial constraints or insufficient food supply
\item Substance use – Use of illegal drugs, misuse of prescription drugs, or problematic use of regulated substances
\item Alcohol use – Excessive consumption of alcohol, binge drinking, or alcohol-related issues affecting daily routines
\item Mental health – Depressive symptoms, anxiety, suicidal ideation, or psychiatric manifestations impacting daily functioning

Examples of the social needs category “mental health” (non‐exhaustive list):
\begin{itemize}
\item Hallucinations (perceiving things that are not really present)
\item Delusions (firmly held false beliefs)
\item Pronounced mood swings (e.g., during manic or depressive episodes)
\item Paranoia or unfounded fears
\item Impaired insight or judgment
\item Bipolar Disorder
\item Schizophrenia
\item Post-Traumatic Stress Disorder (PTSD)
\item Obsessive-Compulsive Disorder (OCD)
\end{itemize}

Examples where the social needs category “mental health” should not be noted:
\begin{itemize}
\item ...
\end{itemize}
\end{itemize}

\item For each identified social need, provide:
\begin{enumerate}[label=\alph*.]
\item A brief description detailing the unique aspects of the social need (specifics about the patient's situation)
\item A classification of the need's status, which must be exactly one of the following: “resolved” – the need existed but has been addressed and is no longer active; “ongoing” – the need continues to be present; “new” – the need has been newly identified in the current encounter
\end{enumerate}

Extract actions corresponding to each identified need. Categorize these actions as:
\begin{enumerate}[label=\alph*.]
\item Previous actions – Completed interventions (these should not be recommendations) such as past referrals, connections, or consults:
\begin{itemize}
\item Description of what was done
\item Start date (if available)
\item Outcome
\end{itemize}
\item Current actions (in progress) – Active interventions (not recommendations) including referrals, connections, or consults that are currently underway:
\begin{itemize}
\item Description of what is being done
\item Start date (if available)
\end{itemize}
\item Planned actions – Scheduled or intended future interventions (again, not recommendations) including those yet to be initiated:
\begin{itemize}
\item Description of what needs to be done
\end{itemize}
\end{enumerate}

When picking out actions, search for interventions such as (this is a partial list):
\begin{itemize}
\item For Mental Health: placement in IMD facilities, LSAT, MHRC, ADU, psychiatric holds/conservatorship
\item For Substance Use: referrals for detox or withdrawal services, residential/outpatient SUD programs
\item For Housing/Discharge: interventions such as Home Health (PT/OT/SLP/RN), shelters, navigation centers, Medical Respite care, Coordinated Entry (CE) assessments, SNF referrals, Board \& Care/ALF/RCFE placements, ARU/ARF, or LTACH care
\end{itemize}

\item Identify the patient's current housing situation according to one of the following exact categories:
\begin{itemize}
\item Rent or Own – The individual has stable, independent housing with legal tenancy rights confirmed via a rental agreement or property ownership. SROs (Single Room Occupancy with shared bathrooms and sometimes shared kitchens) are acceptable if onsite services are not provided.
\item Rent Unit with Support Services Onsite – Housing under a subsidized or supportive model where integrated services (such as case management or mental health support) are available in the same building.
\item Residential Treatment / Jail - will exit homeless – The individual is currently in an institutional setting and is expected to have no housing after discharge or release.
\item Living Outside (Street / Vehicle / Makeshift) –  
  \begin{itemize}
  \item Street – Living outdoors (e.g., on the street, sidewalk, doorway, park, or freeway underpass).
  \item Vehicle – Living in a car, van, bus, truck, RV, or similar vehicle.
  \item Makeshift – Residing in an enclosure or structure not approved for habitation by building or housing codes (e.g., an abandoned building).
  \end{itemize}
\item Temporarily with Friend / Family – An unstable housing situation where the individual is staying with others.
\item Stabilization Room / Hotel – Living in a hotel room without formal tenancy rights.
\item Shelter / Navigation Center – Currently residing in a supervised temporary facility (such as a homeless shelter or mission).
\item Permanent Supportive Housing – Residing within a residential care or board and care facility. SROs are acceptable if there are no onsite services.
\item Residential Treatment / Jail - will exit housed – The individual is in an institutional setting but has arranged for housing upon discharge or release.
\item Unknown
\end{itemize}

\begin{enumerate}[label=\alph*.]
\item Provide a concise description of the patient’s housing situation.
\end{enumerate}

\item For social work consult note types only, determine the reason for the social work consult if it is explicitly stated or if it clearly indicates “Reason for SW…”. If the consult reason is ambiguous or the note does not pertain to a social work consult, do not extract this information.

Examples where the reason for the social worker consult should be recorded:
\begin{itemize}
\item ...
\end{itemize}

Examples where the consult reason should not be annotated:
\begin{itemize}
\item ...
\end{itemize}

\item Identify the reason for the admission and the diagnosis for the current encounter.

\item Outpatient Therapy and Home Health Services – Determine if the patient benefits from at-home services by health workers (or is in a Skilled Nursing Facility) or if a SNF referral has been made (this should not be a recommendation). Examples include services provided by PT/OT/SLP/RN/home health aides or IHSS caregivers (not family members).

Examples where Outpatient Therapy and home health services should be annotated:
\begin{itemize}
\item ...
\end{itemize}

Examples where Outpatient Therapy and home health services should not be noted:
\begin{itemize}
\item ...
\end{itemize}

\item Identify explicit references to the patient receiving or having consults in the following inpatient therapies (not recommendations):
\begin{itemize}
\item Rehab
\item Speech
\item Nutrition
\item Language
\end{itemize}

For each therapy mentioned:
\begin{enumerate}[label=\alph*.]
\item Indicate whether there is a consult for the therapy (yes, no, or unknown)
\item Provide a brief summary of the patient’s progress in that specific therapy
\end{enumerate}

Examples where Inpatient Therapy should be annotated:
\begin{itemize}
\item ...
\end{itemize}

Examples where Inpatient Therapy should not be annotated:
\begin{itemize}
\item ...
\end{itemize}

\item Identify any mentions of the patient’s contacts. For example, any reference to friends, family members, or case workers:
\begin{itemize}
\item ...
\end{itemize}

\item Provide a brief summary of any evidence indicating the patient has few or minimal contacts, or appears socially isolated. Examples include indications such as being widowed, living alone without clear family support, lack of a next of kin, unavailability of emergency contacts, or expressions of feeling alone or unsupported.

Examples where minimal contacts should be annotated:
\begin{itemize}
\item ...
\end{itemize}

\item Identify any difficulties the patient has in performing ADLs (Activities of Daily Living) such as bathing, dressing, toileting, grooming, or eating.
\begin{enumerate}[label=\alph*.]
\item Provide a brief description detailing the issues with ADLs
\end{enumerate}

\item Identify any challenges the patient faces when performing IADLs (Instrumental Activities of Daily Living) which include more complex tasks like managing medications, using transportation, cooking, handling finances, shopping, housekeeping, or managing appointments.
\begin{enumerate}[label=\alph*.]
\item Provide a brief description detailing the issues with IADLs
\end{enumerate}

\item If the patient has a history of substance use or is currently struggling with it, identify the following:
\begin{enumerate}[label=\alph*.]
\item A brief timeline summarizing the patient's substance use history
\item Date of the last substance use (if available)
\item Indicate whether the patient exhibits a problematic pattern of opioid use (including fentanyl or heroin) that results in clinically significant impairment or distress, or if the note mentions consistent long-term opioid use, or explicitly identifies the patient as having OUD (yes, no, unknown)
\item Indicate whether the patient shows a problematic pattern of using amphetamine-type substances, cocaine, methamphetamine, or other stimulants leading to clinically significant impairment or distress, or if the note mentions chronic methamphetamine use or explicitly identifies the condition as StUD (yes, no, unknown)
\item Polysubstance Use Disorder (PSUD) – Note if the patient is explicitly stated to have or to have had PSUD
\item Medication for Opioid Use Disorder (MOUD) – Note if the patient is on methadone, buprenorphine (e.g., Suboxone, Brixadi, Sublocade); indicate with “yes”, “no”, or “unknown”
\end{enumerate}

\item Identify each piece of durable medical equipment (DME) the patient currently utilizes or requires. Examples include items such as mobility devices, wheelchairs, walkers, canes, crutches, electric hospital beds, Hoyer/transfer lifts with slings, bedside commodes, shower chairs/tub benches, raised toilet seats with arms, grab bars, home oxygen concentrators, portable oxygen cylinders, nebulizers, CPAP/BiPAP units, portable suction machines, negative-pressure wound-vac pumps, ambulatory/elastomeric infusion pumps, IV poles (or similar devices)
\begin{enumerate}[label=\alph*.]
\item Provide a concise explanation for the usage of each piece of DME
\end{enumerate}

\item Provide a short summary of any recurring patterns in the patient’s ED visits

\item Provide a brief summary of any recurring patterns in the patient’s psychiatric visits. Include whether the patient typically presents as “acute”, as an “acute diversion” (e.g., ADU), or for “respite” care

\item Provide a summary of the support system available to the patient at home

\item Determine whether this patient is high priority (yes, no, unknown) and justify your reasoning.

High-priority patients include those with at least one of the following:
\begin{itemize}
\item Homeless and discharge ready – Patients who are medically stable for discharge but lack stable housing. This includes shelter residents, those living in unapproved dwellings, or those unable to return to previous housing.
\item Jane/John Doe – Patients whose identity has not been verified or established. This includes unconscious patients lacking identification or those unable to communicate accurate identifying information.
\item Suffering from Abuse/Neglect/Self-Neglect/Safety Concerns – Cases where there is evidence of physical, emotional, sexual, financial, or psychological harm, inadequate caregiver attention resulting in harm or risk, self-neglect leading to failure in meeting basic needs, or overall conditions that compromise safety upon discharge or at home.
\item Need family located – Situations that require identifying and contacting family members or next of kin for purposes such as medical decision-making, discharge planning, or critical medical updates.
\item Death and Dying – Any mention that the patient has expired, is actively dying, or that discussions of end-of-life planning (e.g., hospice enrollment, DNR orders, comfort care) are underway.
\item Major Life Changing Event – An abrupt occurrence that permanently alters the patient’s daily functioning or psychosocial status and necessitates new or long-term planning. This may include a newly acquired permanent disability, life-changing diagnosis, loss of a primary caregiver, or housing explicitly described as permanently altered (e.g., limb amputation, sudden paralysis)
\end{itemize}
\end{enumerate}

\textit{Note: In-context learning examples have been excluded to protect patient health information privacy}

\subsection{Example LLM-as-a-judge Prompt}

Your task is to assess answers for a question based on provided patient note(s) and provide your reasoning. Evaluate each answer using the three criteria below:

\begin{enumerate}
    \item Accuracy - Does the answer correctly and comprehensively address all the questions, using precise and factual clinical information from the source note(s)?
    \begin{itemize}[leftmargin=*]
        \item ``Yes'' if: The answer clearly addresses all aspects of the question, is factually accurate, complete, and fully supported by information from the patient notes. Note: If the answer states ``no information found'' and there truly is no relevant information in the notes, this is considered accurate for all questions
        \item ``No'' if: The answer fails to respond to the core prompt, contains significant factual errors, omits critical information, or misrepresents details from the notes.
    \end{itemize}
    
    \item Relevance - How relevant is all the cited content from the clinical note(s) for the correct answer?
    \begin{itemize}[leftmargin=*]
        \item Score 1 (Low): The cited information is not relevant to the correct answer.
        \item Score 2 (Moderate): Some of the cited information is partially relevant to the correct answer.
        \item Score 3 (High): All the cited information is fully relevant and directly supports the correct answer.
    \end{itemize}
    
    \item Fluency - How well is the answer presented in terms of clarity, structure, conciseness, and use of appropriate, professional clinical language?
    \begin{itemize}[leftmargin=*]
        \item Score 1 (Low): The text is disorganized, confusing, uses inappropriate terminology, or is excessively wordy, making it difficult to understand.
        \item Score 2 (Moderate): The text is generally clear but may have some issues with structure, conciseness, or professional language.
        \item Score 3 (High): The text is well-structured, concise, easy-to-understand, and uses appropriate professional language for clinical documentation.
    \end{itemize}
\end{enumerate}

\textit{Patient Note(s):}

... \\

\textit{Question}: Identify any inability to perform ADLs (Activities of Daily Living): basic self care tasks bathing, dressing, toileting, grooming, eating.  a. A brief description of the ADLs
\\

% \textit{Answer}

% The note mentions the patient needs maximum assistance with all ADLs

For each criterion, think through your reasoning systematically before providing your final assessment.

\section{Additional Results}

\begin{table*}[ht]
\centering
\begin{tabular}{p{3.2cm}|>{\centering\arraybackslash}p{5cm}|>{\centering\arraybackslash}p{1.2cm}|>{\centering\arraybackslash}p{1.2cm}}
 & Concordance difference (95\% CI) & t-statistic & p-value \\
\midrule
Alcohol Use & 0.26 (0.174, 0.351) & 5.91 & $<.001$ \\
Medical Equipment & 0.08 (0.026, 0.135) & 2.94 & 0.004 \\
Food Insecurity & -0.01 (-0.030, 0.010) & -1.00 & 0.320 \\
Housing & 0.18 (0.090, 0.274) & 3.93 & $< .001$ \\
Immigration & 0.06 (0.013, 0.108) & 2.51 & 0.014 \\
Mental Health & 0.38 (0.060, 0.243) & 3.28 & $<.001$ \\
Patient Contacts & 0.21 (0.125, 0.299) & 4.85 & $< .001$ \\
Safety & 0.06 (-0.029, 0.150) & 1.35 & 0.181 \\
Tobacco Use & 0.26 (0.170, 0.355) & 5.62 & $< .001$ \\
Reason for Consult  & 0.38 (0.321, 0.527) & 8.17 & $< .001$ \\
Substance Use  & 0.20 (0.099, 0.305) & 3.91 & $< .001$ \\
Outpatient Therapy & 0.39 (0.288, 0.500) & 7.38 & $< .001$ \\
Opiod Use & 0.82 (-0.033, 0.154) & 1.28 & 0.202 \\
\bottomrule
\end{tabular}
\caption{
Difference in concordance between LLM application's extractions and LLM judge, where positive differences means an increase in concordance from \promptone{} to \promptfour{}}
\label{app:cis}
\end{table*}

% \subsection{Human Validation Results}

\begin{table}[htbp]
\label{app:human_annotation_res}
\centering

\resizebox{\textwidth}{!}{
\begin{tabular}{l|ccc|ccc}
\toprule
 & \multicolumn{3}{c|}{\textbf{Cohen's Kappa (95\% CI)}} & \multicolumn{3}{c}{\textbf{Gwet's AC1 (95\% CI)}} \\
\midrule
\textbf{Category} & \textbf{Inter-annotator} & \textbf{LLM app vs annotator} & \textbf{LLM judge vs annotator} & \textbf{Inter-annotator} & \textbf{LLM app vs annotator} & \textbf{LLM judge vs annotator} \\
\midrule
Alcohol Use & 0.00 (0.00, 0.00) & 0.64 (0.11, 0.90) & 0.03 (-0.06, 0.15) & 0.93 (0.73, 1.00) & 0.97 (0.94, 0.99) & 0.85 (0.76, 0.92) \\
Food Insecurity & 1.00 (1.00, 1.00) & 0.24 (-0.01, 0.43) & 0.38 (-0.03, 0.93) & 1.00 (1.00, 1.00) & 0.94 (0.88, 0.99) & 0.98 (0.95, 1.00) \\
Housing & 0.76 (0.50, 1.00) & 0.41 (0.13, 0.67) & 0.69 (0.39, 0.90) & 0.75 (0.45, 1.00) & 0.80 (0.67, 0.90) & 0.90 (0.82, 0.97) \\
Immigration & -- & 0.00 (0.00, 0.00) & -- & 1.00 (1.00, 1.00) & 0.98 (0.94, 1.00) & 1.00 (1.00, 1.00) \\
Medical Equipment & 1.00 (1.00, 1.00) & 0.81 (0.65, 0.94) & 0.85 (0.69, 0.96) & 1.00 (1.00, 1.00) & 0.89 (0.79, 0.97) & 0.91 (0.82, 0.98) \\
Mental Health & -0.19 (-0.44, 0.00) & 0.38 (0.16, 0.58) & 0.42 (0.20, 0.69) & 0.44 (-0.41, 0.93) & 0.70 (0.54, 0.82) & 0.81 (0.69, 0.91) \\
Opioid Use & 0.40 (0.40, 0.40) & 0.83 (0.51, 1.00) & 0.82 (0.54, 1.00) & 0.92 (0.68, 1.00) & 0.98 (0.95, 1.00) & 0.98 (0.96, 1.00) \\
Outpatient Therapy & 0.60 (0.00, 1.00) & 0.65 (0.45, 0.83) & 0.65 (0.45, 0.82) & 0.82 (0.34, 1.00) & 0.82 (0.70, 0.93) & 0.76 (0.62, 0.89) \\
Patient Contacts & 0.30 (0.00, 0.50) & 0.39 (0.23, 0.57) & 0.60 (0.36, 0.79) & 0.84 (0.40, 1.00) & 0.49 (0.30, 0.67) & 0.80 (0.67, 0.90) \\
Activities of Daily Living & 0.60 (0.00, 1.00) & 0.86 (0.71, 0.97) & 0.79 (0.59, 0.95) & 0.82 (0.34, 1.00) & 0.93 (0.85, 0.98) & 0.92 (0.84, 0.98) \\
Reason for Admission & 0.75 (0.16, 1.00) & 0.67 (0.50, 0.81) & 0.69 (0.54, 0.85) & 0.75 (0.26, 1.00) & 0.67 (0.51, 0.82) & 0.69 (0.53, 0.86) \\
Reason for Consult & 0.57 (0.29, 0.88) & 0.36 (0.21, 0.52) & 0.48 (0.28, 0.67) & 0.67 (0.29, 1.00) & 0.31 (0.12, 0.52) & 0.65 (0.50, 0.80) \\
Safety & 0.17 (-0.13, 0.50) & 0.39 (0.10, 0.69) & 0.59 (0.19, 0.91) & 0.73 (0.18, 1.00) & 0.81 (0.70, 0.90) & 0.89 (0.81, 0.96) \\
Substance Use & 0.80 (0.50, 1.00) & 0.51 (0.16, 0.77) & 0.49 (0.14, 0.77) & 0.91 (0.67, 1.00) & 0.83 (0.72, 0.92) & 0.81 (0.69, 0.90) \\
Tobacco Use & 0.36 (0.00, 0.50) & 0.40 (0.29, 0.48) & 0.62 (0.20, 0.88) & 0.82 (0.36, 1.00) & 0.94 (0.88, 0.99) & 0.86 (0.77, 0.93) \\
\midrule
\textbf{Overall} & \textbf{0.62 (0.48, 0.74)} & \textbf{0.60 (0.56, 0.65)} & \textbf{0.67 (0.63, 0.71)} & \textbf{0.84 (0.78, 0.90)} & \textbf{0.85 (0.83, 0.86)} & \textbf{0.89 (0.88, 0.91)} \\
\bottomrule
\end{tabular}
}
\label{tab:agreement_stats}
\caption{Cohen's Kappa and Gwet's AC1 Agreement Statistics}
\end{table}

\end{document}